\documentclass[fleqn,usenatbib]{mnras}

\usepackage{newtxtext,newtxmath}

\usepackage[T1]{fontenc}

\DeclareRobustCommand{\VAN}[3]{#2}
\let\VANthebibliography\thebibliography
\def\thebibliography{\DeclareRobustCommand{\VAN}[3]{##3}\VANthebibliography}

\usepackage{graphicx}	
\usepackage{amsmath}	

\title[The Iron Cross-Correlation Signal of WASP-76b]{Decomposing the Iron Cross-Correlation Signal of the Ultra-Hot Jupiter WASP-76b in Transmission using 3D Monte-Carlo Radiative Transfer}

\author[Joost P. Wardenier et al.]{
Joost P. Wardenier,$^{1}$\thanks{E-mail: joost.wardenier@physics.ox.ac.uk}
Vivien Parmentier,$^{1}$
Elspeth K.H. Lee,$^{2}$
Michael R. Line,$^{3}$
Ehsan Gharib-Nezhad$^{4}$
\\
$^{1}$Department of Physics (Atmospheric, Oceanic and Planetary Physics), University of Oxford, Oxford, OX1 3PU, UK\\
$^{2}$Center for Space and Habitability, University of Bern, Gesellschaftsstrasse 6, CH-3012 Bern, Switzerland\\
$^{3}$School of Earth \& Space Exploration, Arizona State University, Tempe AZ 85287, USA \\
$^{4}$NASA Ames Research Center, Moffett Field, CA 94035, USA
}

\date{Accepted 2021 June 17. Received 2021 May 20; in original form 2021 April 7}

\pubyear{2015}

\begin{document}
\label{firstpage}
\pagerange{\pageref{firstpage}--\pageref{lastpage}}
\maketitle

\begin{abstract}
Ultra-hot Jupiters are tidally locked gas giants with dayside temperatures high enough to dissociate hydrogen and other molecules. Their atmospheres are vastly non-uniform in terms of chemistry, temperature and dynamics, and this makes their high-resolution transmission spectra and cross-correlation signal difficult to interpret. In this work, we use the SPARC/MITgcm global circulation model to simulate the atmosphere of the ultra-hot Jupiter WASP-76b under different conditions, such as atmospheric drag and the absence of TiO and VO. We then employ a 3D Monte-Carlo radiative transfer code, \textsc{hires-mcrt}, to self-consistently model high-resolution transmission spectra with iron (Fe \textsc{i}) lines at different phases during the transit. To untangle the structure of the resulting cross-correlation map, we decompose the limb of the planet into four sectors, and we analyse each of their contributions separately. Our experiments demonstrate that the cross-correlation signal of an ultra-hot Jupiter is primarily driven by its temperature structure, rotation and dynamics, while being less sensitive to the precise distribution of iron across the atmosphere. We also show that the previously published iron signal of WASP-76b can be reproduced by a model featuring iron condensation on the leading limb. Alternatively, the signal may be explained by a substantial temperature asymmetry between the trailing and leading limb, where iron condensation is not strictly required to match the data. Finally, we compute the $K_{\text{p}}$--$V_{\text{sys}}$ maps of the simulated WASP-76b atmospheres, and we show that rotation and dynamics can lead to multiple peaks that are displaced from zero in the planetary rest frame.
\end{abstract}




\section{Introduction}
\label{sect:intro}

One of the main goals associated with the characterization of exoplanets is to quantify the abundances of chemical species in their atmospheres. Nonetheless, the inherent 3D structure of exoplanets complicates this endeavour. A suite of studies in low spectral resolution have demonstrated how interpreting observations with 1D models leads to \mbox{\emph{biased}} inferences, both in the case of transmission spectra (\citealt{Line2016a,MacDonald2017,Caldas,Pluriel2020,MacDonald2020,Lacy2020}) and emission spectra (\citealt{Feng2016,Blecic2017,Taylor2020}). In high resolution, where individual spectral lines are resolved, the problem becomes even more intricate. This is because line shapes, depths and positions depend profoundly on the 3D thermal structure and chemical composition of the planet, as well as its wind profile and rotation (\citealt{Miller-RicciKempton2012, Showman2013, M-RKempton2014, Rauscher2014,Zhang2017,Flowers2019,Harada2019,Seidel2019a,Beltz2020,Keles2021}). Hence, comparing observational data to the wrong models may lead to wrong conclusions about the nature of the atmosphere. In fact, to extract the correct physical information, it is crucial to understand how spatial inhomogeneities across the atmosphere show up in transmission and emission spectra, and how these can be accounted for in models. 

\begin{figure*}
\centering

\vspace{-10pt}

\makebox[\textwidth][c]{ \includegraphics[width=1.1\textwidth]{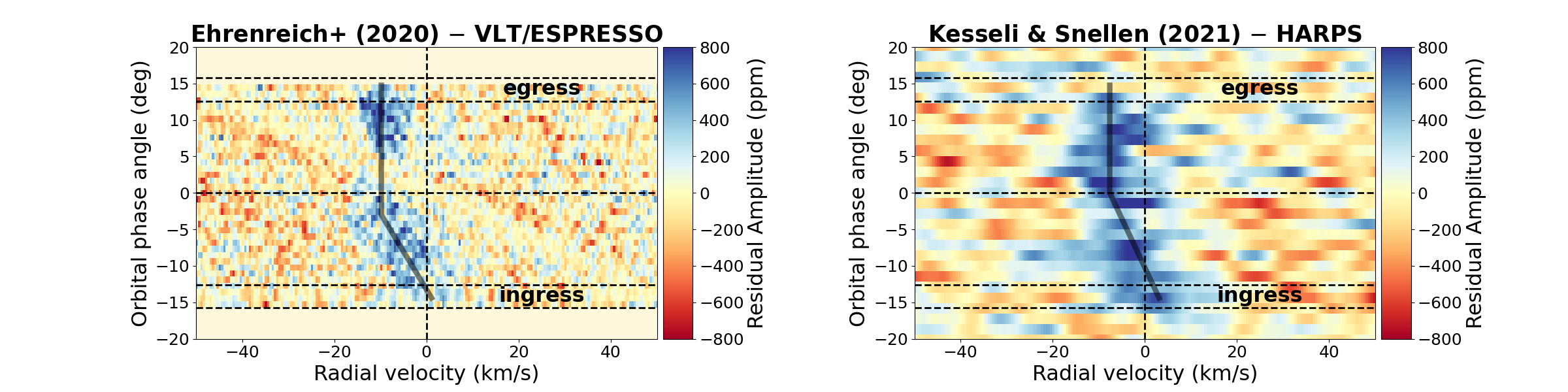}}

\vspace{-5pt}

\caption{The asymmetric iron absorption signal of WASP-76b, as measured by \citet{Ehrenreich2020} and \citet{Kesseli2021}. The observations were transformed to the planetary rest frame, to exclude contributions from orbital and systemic motion. The maps were obtained by cross-correlating the planetary spectra with templates containing iron lines. The signals indicate the average radial velocity by which gaseous iron (Fe \textsc{i}) in the observable atmosphere of WASP-76b is moving with respect to the rest frame. The solid black lines indicate the overall trends, but are by no means a fit to the data. The signal from \citet{Ehrenreich2020} fades between 0 and 5 degrees, due to an overlap between the Doppler shadow of the star and the planetary signal.}
\label{fig:ehrenreich_rv}
\end{figure*}


Situated in the closest vicinity of their host stars ($<$0.05 AU) and having no counterparts in our Solar system, ultra-hot Jupiters (\citealt{Bell2018, Parmentier2018, Arcangeli2018}) are ideal testbeds for studying the impact of 3D effects on high-resolution spectra. There are two important reasons for this. Firstly, ultra-hot Jupiters are \emph{accessible} objects to observe. Their short orbital periods (1--2 days) and hot, extended atmospheres make them perfect targets for transmission spectroscopy (\citealt{Hoeijmakers2019, VonEssen2019, Ehrenreich2020, Borsa2021}), emission spectroscopy (\citealt{Evans2017, Arcangeli2018, Mikal-Evans2020}) and phase-curve studies (\citealt{Zhang2018, Mansfield2020, Bourrier2020b}). Secondly, ultra-hot Jupiters display extreme \emph{variations} across their atmospheres, because they are expected to become tidally locked soon after their formation (\citealt{Rasio1996, Showman2002}). As a result, their atmospheres virtually consist two different worlds: a permanently irradiated dayside and a permanently dark nightside. The scorching, cloud-free dayside ($T\gtrsim2500$ K) nearly resembles a stellar photosphere, where most molecules are dissociated\footnote{By definition, ultra-hot Jupiters are planets hot enough to dissociate molecules (e.g. water and hydrogen) on their daysides. For a planet like WASP-76b this happens at dayside temperatures above $\sim$$2400$ K. However, on higher-gravity objects, such as brown dwarfs, dissociation requires \emph{higher} temperatures (e.g., Figure 13 in \citealt{Parmentier2018}).} and metals become ionised (\citealt{Parmentier2018, Hoeijmakers2019}). On the other hand, the nightside is substantially cooler ($T\lesssim1000$ K) and may even serve as a stage for cloud formation (\citealt{Helling2019, Ehrenreich2020}). Ultra-hot Jupiters also exhibit large differences in their thermal structures: the dayside is expected to show strong thermal inversions (\citealt{Haynes2015, Evans2017, Kreidberg2018,Yan2020, Pino2020}), whereas nightside temperatures are expected to monotonically decrease with altitude. Furthermore, ultra-hot Jupiters feature strong winds in the order of 1--10 km/s (\citealt{Tan2019}), which arise as a result of the continuous day-night forcing. Many observational studies have measured Doppler shifts due to winds on ultra-hot Jupiters (\citealt{Casasayas-Barris2019, Ehrenreich2020, Hoeijmakers2020, Bourrier2020, Cabot2020, Gibson2020, Nugroho2020, Stangret2020, Tabernero2020, Borsa2021, Kesseli2021, Rainer2021}), yet inferring the underlying 3D circulation pattern is a formidable challenge.




Many of the observational insights regarding the structure, composition and dynamics of (ultra-)hot Jupiters have been gained from ground-based high-resolution spectroscopy (HRS, \citealt{Snellen2010, Brogi2012, Birkby2018}). In high resolution, individual spectral lines are resolved, and this has unique advantages over low-resolution observations (for methods to combine high- and low-resolution data, see \citealt{Brogi2017,Pino2018a,Gandhi2019b,Brogi2019}). Firstly, the planet signal can be separated from stellar and telluric contributions thanks to the orbital motion of the planet, which (periodically) Doppler shifts its spectrum. Secondly, HRS is sensitive to the unique spectral fingerprint of atoms and molecules. This allows for the unambiguous detection of chemical species in the atmosphere of a planet, generally in combination with the cross-correlation technique (\citealt{Snellen2010, Brogi2012, Brogi2014, Brogi2016, Rodler2013, DeKok2013, Birkby2013, Birkby2017, Wyttenbach2015, Schwarz2016, Nugroho2017, Nugroho2020, Hawker2018, Hoeijmakers2018, Hoeijmakers2019, Cauley2019, Alonso-Floriano2019a, Yan2019, Gibson2020, Giacobbe2021}). The benefit of this approach is that the information from many lines can be combined simultaneously, which boosts the signal-to-noise ratio of the detection (\citealt{Snellen2015}). Thirdly, HRS is sensitive to Doppler shifts and Doppler broadening caused by atmospheric dynamics and planetary rotation. Hence, the observed line shapes, depths and positions can be used to constrain wind patterns (\citealt{Snellen2010,Louden2015,Salz2018,Flowers2019, Seidel2019a, Bourrier2020}) and, in some cases, rotation rates (\citealt{Snellen2014, Brogi2016, Schwarz2016}). Finally, HRS is capable of probing the outer helium envelopes that some planets exhibit, offering a window into physical processes such as atmospheric escape (\citealt{Nortmann2018, Allart2018,Yan2018}).

The aim of this work is to better understand how the 3D structure and dynamics of ultra-hot Jupiters impact their high-resolution transmission spectra, as well as the resulting cross-correlation signal. To this end, we feed the outputs of a 3D, non-grey global circulation model (GCM) into a Monte-Carlo radiative transfer code that is able to compute Doppler-shifted spectra at different stages during the transit. We apply our framework to the transiting ultra-hot Jupiter WASP-76b, for which \citet{Ehrenreich2020} and \citet{Kesseli2021} independently reported an asymmetric iron absorption signal (see Figure \ref{fig:ehrenreich_rv}) -- an observation that could testify to 3D effects in the planet's atmosphere. In Section \ref{sect:w76}, we briefly revisit the observation and discuss its current interpretation. In Section \ref{sect:methods}, we outline how we generate our atmospheric structures with the GCM and how we use the Monte-Carlo code to compute transmission spectra. Additionally, we elaborate on the methods used to compute cross-correlation maps. The atmospheric structures and a number of spectra are presented in Section \ref{s:spec_and_struc}. Subsequently, in Sections \ref{s:results_weak_drag} and \ref{s:results_all_gcm}, we present the cross-correlation maps and we attempt to link their behaviour to the physics of the atmospheres. The corresponding \mbox{$K_{\text{p}}$--$V_{\text{sys}}$} maps are discussed in Section \ref{sect:kpvsys}. Finally, Section \ref{sec:conclusion} provides a summary and conclusion.

\begin{figure}
\centering
\includegraphics[width=0.48\textwidth]{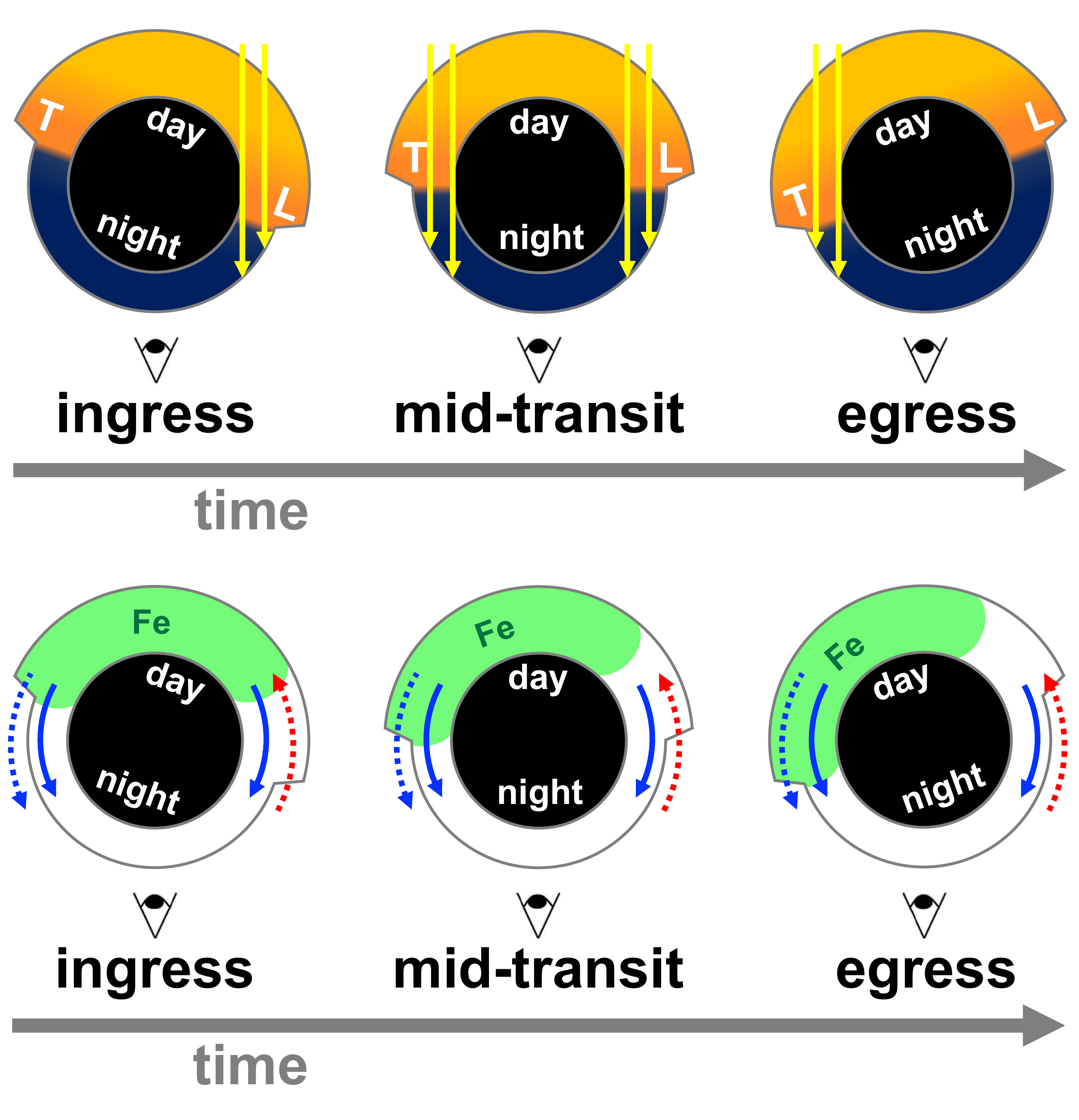}
\vspace{-10pt}
\caption{The equatorial plane of WASP-76b. \textbf{Top row:} The orientation of the planet at three stages during the transit. The yellow region denotes the hot dayside atmosphere and the blue region the cooler nightside. The outer contour of the images represents an isobar, which lies further from the center of the planet on the (puffy) dayside than on the (compact) nightside. The pairs of yellow arrows show stellar light rays crossing the atmosphere. The trailing and leading limb of the planet are labelled \texttt{T} and \texttt{L}, respectively. \textbf{Bottom row:} The iron (Fe \textsc{i}) distribution of WASP-76b as proposed by \citet{Ehrenreich2020}. The green region on the dayside is abundant in iron, while the white region is depleted. Moreover, the solid arrows show the direction of day-to-night winds, while the dotted arrows indicate the direction of planetary rotation. Note that angles and sizes in the figure have been exaggerated for visualisation purposes.}
\label{fig:day_night_schematic}
\end{figure}

\section{The iron signal of WASP-76b in High Spectral Resolution}
\label{sect:w76}

WASP-76b (\citealt{West2016}) is an archetypal ultra-hot Jupiter. With a semi-major axis of 0.033 AU, the planet orbits an F7-type main sequence star in 1.81 days. Recently, \citet{Ehrenreich2020} observed WASP-76b's transit with VLT/ESPRESSO (\mbox{$R$ = 138,000}) in both high \emph{spectral} and \emph{temporal} resolution, and claimed evidence for neutral iron (Fe \textsc{i}) condensation on its nightside. Transit observations provide valuable insights into the non-uniformity of ultra-hot Jupiters, because they probe the terminator region of the atmosphere. The terminator region constitutes the interface between the dayside and the nightside of the planet, and can thus be expected to feature extreme thermal and chemical gradients. Because WASP-76b resides so close to its host star, the planet rotates by no less than 30 degrees during its transit. In combination with the gradients throughout the atmosphere, this large angle must inevitably lead to changes in the transmission spectrum with time. Such changes can only be interpreted with 3D models.

\citet{Ehrenreich2020} indeed reported a time-dependence in the transmission spectrum of WASP-76b. Across the wavelength range covered by VLT/ESPRESSO (0.38--0.79 micron), gaseous iron is responsible for most of the absorption lines contained in the spectrum. In high spectral resolution, these lines are swamped by noise, so the spectrum must be multiplied with a template (e.g. a stellar mask or a modelled transmission spectrum) for the planet signal to be recovered. By shifting the template along the wavelength axis, and performing the multiplication many times, one obtains a cross-correlation function (CCF, see Section \ref{ss:methods_ccf}). The peak position of the CCF reveals the average radial velocity (RV) by which the iron in the observable atmosphere is moving with respect to the observer. Figure \ref{fig:ehrenreich_rv} shows the iron absorption signal that \citet{Ehrenreich2020} measured, along with the signal reported by \citet{Kesseli2021}, who applied the same technique to HARPS data (\mbox{$R$ = 115,000}; \mbox{0.38--0.69} micron) to independently confirm the time-dependence of the observation. It should be noted that the signals in Figure \ref{fig:ehrenreich_rv} pertain to the planetary rest frame, which moves with the planet as it orbits the star. In this frame, the CCF of a static atmosphere without winds and rotation will always have a peak at 0 km/s, because there is no moving material to Doppler shift the absorption lines. Although the measurements from \citet{Ehrenreich2020} and \citet{Kesseli2021} may differ slightly in terms of RV values, their qualitative behaviour is very similar. As illustrated in Figure \ref{fig:ehrenreich_rv}, the iron signal of WASP-76b lies close to 0 km/s at ingress, but the spectrum becomes progressively more blueshifted in the first half of the transit. Around mid-transit, the signal features a ``kink'', after which the RV remains more or less constant. At egress, the \citet{Ehrenreich2020} signal is located at $-$11 km/s, while that of \citet{Kesseli2021} ends up near $-8$ km/s.

To explain the behaviour of the iron signal, \citet{Ehrenreich2020} proposed an asymmetry between the trailing and leading limb\footnote{In this work, the \emph{trailing limb} will be synonymous for the evening terminator (or east limb). It is the part of the atmosphere that rotates away from the star and towards the observer. Likewise, we use the term \emph{leading limb} to refer to the morning terminator (or west limb), which rotates towards the star and away from the observer (See Figure \ref{fig:day_night_schematic}). We are aware that some studies use more stringent definitions for the \emph{limb} and the \emph{terminator} (e.g., footnote 1 in \citealt{Caldas}).} of the planet, where the leading limb is largely depleted of gaseous iron, possibly due to cloud formation. Figure \ref{fig:day_night_schematic} summarises what happens in this scenario. At ingress, the stellar light only penetrates the leading limb. Gaseous iron is carried by \emph{blueshifting} day-to-night winds that counteract the \emph{redshifting} planetary rotation. These contributions add up to a net Doppler shift of roughly zero. As the trailing limb moves in front of the star, the signal becomes more blueshifted. This is because the day-to-night flow and planetary rotation both move iron towards the observer on this side of the planet. In the meanwhile, the signal from the leading limb becomes weaker as the amount of iron along the line of sight decreases. After mid-transit, the iron on the leading limb has rotated out of view completely and the signal dominated by the trailing limb.

WASP-76b is not the only ultra-hot Jupiter for which \emph{time-varying} iron absorption has been reported in transmission. \citet{Borsa2021} observed WASP-121b with VLT/ESPRESSO and found a signal very similar to that of \cite{Ehrenreich2020} and \cite{Kesseli2021}, with the iron lines shifting to bluer wavelengths during the transit. Earlier on, WASP-121b observations by \citet{Bourrier2020} had already demonstrated that the average blueshift of iron is larger in the second half of the transit than in the first half \mbox{($-6.6$ vs.} \mbox{$-3.8$ km/s}). \citet{Hoeijmakers2020} studied iron absorption in the atmosphere of MASCARA-2b/KELT-20b and measured a time variability in the signal \emph{strength} of singly-ionised iron \mbox{(Fe \textsc{ii})}. Furthermore, they found that the Fe \textsc{i} and Fe \textsc{ii} signals were blueshifted by different amounts, in agreement with earlier observations by \citet{Stangret2020}. This result could indicate that the absorption lines of both species probe different layers of the atmosphere, each with its own wind speed. In the meanwhile, \citet{Nugroho2020} reported possible signatures of atmospheric dynamics and 3D effects based on the $K_{\text{p}}$--$V_{\text{sys}}$ map (see Section \ref{ss:methods_kpvsys}) of MASCARA-2b/KELT-20b. \citet{Rainer2021} also constructed $K_{\text{p}}$--$V_{\text{sys}}$ maps and even found evidence for changes  \emph{between} different transits of the planet. As for KELT-9b, the hottest ultra-hot Jupiter known to date, \citet{Hoeijmakers2019} did not find a significant Doppler shift in the absorption signal of iron. \citet{Cauley2019}, however, did report time variability in the \emph{width} of the absorption lines. 

Although these observations pertain to just a small set of ultra-hot Jupiters, they already highlight the plethora of ways in which 3D~effects can manifest themselves in the data.



\section{Methods}
\label{sect:methods}

\subsection{Global Circulation Model}
\label{ss:methods_gcm}

We use the SPARC/MITgcm global circulation model (\citealt{Showman2009}) to simulate the atmosphere of WASP-76b under the assumption of tidal locking. The model solves the primitive equations on a cubed-sphere grid and combines 3D atmospheric dynamics with non-grey radiative transfer. The SPARC/MITgcm has been applied to a variety of hot Jupiters (\citealt{Showman2009, Showman2015, Fortney2010, Parmentier2013, Parmentier2016, Parmentier2020, Kataria2015, Kataria2016, Lewis2017, Steinrueck2019, Steinrueck2020}), as well as a number of ultra-hot Jupiters (\citealt{Parmentier2018, Arcangeli2019, Pluriel2020}). To date, however, the only \emph{high-resolution} study that made use of the SPARC/MITgcm was performed by \citet{Showman2013}. We note that \citet{Fortney2010} also calculated transmission spectra from 3D atmospheric models, but their work exclusively focused on low resolution.

\subsubsection{Model Setup}

The setup of our simulations is fully identical to that of \citet{Parmentier2018} $-$ see their Section 2 for more details. The atmosphere has 53 vertical layers, over which the pressure decreases from 200 bar to 2 $\mu$bar. We use a grid resolution of C32, roughly equivalent to 128 cells in longitude and 64 cells in latitude. The models are evolved for 300 days, with time steps of 25 seconds. The final temperature profiles and wind maps are obtained by averaging over the last 100 days of the simulation. Also, we note that the SPARC/MITgcm does not have a self-consistent cloud formation scheme. Instead, it uses equilibrium condensation and gas-phase chemistry to compute the abundances of gaseous species. The effects of iron-cloud formation are mimicked through ``rainout'', where we consider all the iron-bearing gases and condensates from \citet{Visscher2010}. Rainout means that a particular fraction of the gaseous iron is simply removed from cells where the temperature lies below the (pressure-dependent) condensation temperature of a particular iron-bearing condensate.

We model the atmosphere of WASP-76b under four different conditions, summarised in Table \ref{tab:gcm_models}. All models have the same (solar) abundances, but they differ in terms of their drag timescale $\tau_{\rm drag}$. This is the time it takes for winds to lose a significant fraction of their kinetic energy as a result of drag forces (\citealt{Showman2013}, \citealt{Parmentier2017}). These drag forces could stem from processes such as turbulent mixing (\citealt{Li2010}), Lorentz-force braking of ionised winds in the planet's magnetic field (\citealt{Perna2010}) and Ohmic dissipation (\citealt{Perna2010a}). In the absence of drag forces ($\tau_{\text{drag}} \rightarrow \infty$), (ultra-)hot Jupiters develop a superrotating jet around their equator. However, when the drag timescale becomes shorter, winds are slowed down and jet formation is suppressed (\citealt{Komacek2016, Komacek2017}). In this regime, the atmospheric circulation pattern is dominated by day-to-night winds across the entire terminator. We consider atmospheres with $\tau_{\text{drag}}$ values of $\infty$ (no drag), \mbox{$10^5$ s} (weak drag) and \mbox{$10^4$ s} (strong drag), respectively.

\begin{table}
\centering
\caption{Overview of the four WASP-76b scenarios simulated with the SPARC/MITgcm.}
\begin{tabular}{l|c|c}
\hline
\textbf{Model} & \textbf{Drag timescale} & \textbf{TiO and VO opacities} \\ \hline
No drag & $\infty$ & \checkmark \\
Weak drag & $10^5$ s & \checkmark \\
Strong drag & $10^4$ s & \checkmark \\
No drag w/o TiO$+$VO & $\infty$ & $\times$ \\
\hline
\end{tabular}
\label{tab:gcm_models}
\end{table}

Besides these three models, we also study a drag-free atmosphere where the opacities of TiO and VO are assumed to be zero, both in the GCM and in the post-processing (no drag w/o TiO+VO). Such a model mimics the effects of a ``cold trap'' (\citealt{Spiegel2009, Parmentier2013,Beatty2017}), where the nightside condensation of TiO and VO leads to cloud particles that settle gravitationally. If the vertical mixing in the atmosphere is not strong enough to carry these particles aloft again, the upper atmosphere will become depleted of TiO and VO, both on the dayside and the nightside of the planet. Normally, TiO and VO are important optical absorbers and act as thermal-inversion agents in the atmospheres of ultra-hot Jupiters (\citealt{Hubeny2003}, \citealt{Fortney2008}, \citealt{Gandhi2019}). Hence, the presence of a cold trap should change the radiative feedbacks in the atmopshere, resulting in a different temperature structure and wind profile.

In this work, we do not examine the effects of departures from solar-composition abundances. As will be shown later in this manuscript, our results are mainly driven by asymmetries between the trailing and the leading limb of the planet, and we do not expect these to change substantially as a function of metallicity or C/O ratio. A higher metallicity would shift the whole temperature structure of the atmosphere to lower pressures, while increasing the day-night contrast (\citealt{Kataria2015}). As for C/O, only \citet{Mendonca2018} studied the effect of the C/O ratio on the atmospheric circulation. We expect that small C/O variations will have a similar impact as changes in the metallicity, because for C/O$<$1 the only optically important species to become more or less abundant is water (the amount of CO will also change, but is less important to the radiative transfer). To our knowledge, highly non-solar metallicities or C/O ratios larger than unity have not been unambiguously detected in exoplanet atmospheres, so we propose to leave the exploration of this part of parameter space for later work.

\subsubsection{A Note on Previous Studies}

With regards to previous 3D modelling studies, it should be noted that the majority of recent works (\citealt{Miller-RicciKempton2012, Zhang2017, Flowers2019, Harada2019, Beltz2020}) made use of the GCM from \citet{Rauscher2012}, which employs double-gray radiative transfer. This means that the GCM uses one opacity value $\kappa_{\textsc{vis}}$ for shortwave (optical) radiation and another value $\kappa_{\textsc{ir}}$ for longwave (infrared) radiation -- thereby ignoring the more intricate wavelength-dependence of the atmospheric opacities. One of the downsides of this approach is that the model tends to produce isothermal temperature profiles above the photosphere (\citealt{Parmentier2014}), where the atmosphere neither absorbs nor emits a significant amount of radiation. That said, it is hard to judge to what extent high-resolution transmission spectra ``suffer'' from the double-gray approximation. \citet{M-RKempton2014} demonstrated for a suite of hot Jupiters that absorption features are hardly sensitive to the precise \emph{vertical} shape (e.g. inversions) of the temperature structure, which suggests that isothermal profiles may not be problematic in transmission\footnote{The vertical shape of the temperature structure does have a big impact on high-resolution \emph{emission} spectra. For instance, when the atmosphere is isothermal, the emission spectrum will have no features because the core-to-continuum contrast is zero.}, as long as \emph{longitudinal} variations are properly captured. 

The SPARC/MITgcm, employed in this work, uses non-grey radiative transfer to evaluate its heating and cooling rates, resulting in quantitatively different temperature structures and perhaps different 3D wind profiles (\citealt{Showman2013}). However, no detailed comparison between the SPARC/MITgcm and the GCM from \citet{Rauscher2012} has ever been carried out. Also, there will be more differences between both models besides their radiative-transfer treatment, and this makes it hard to interpret any disagreement between their outputs. For example, \citet{Miller-RicciKempton2012} and \citet{Showman2013} both performed simulations of HD 209458b and found different wind speeds in cases where they did not include drag forces. To our knowledge, the precise origin of the discrepancies has never been investigated.


\subsection{Mapping GCM Outputs onto a Spherical Grid}
\label{ss:methods_spherical_grid}

The GCM uses pressure as a vertical coordinate. However, in the context of optical depths and transit spectra, we are actually interested in physical distances along the line of sight. Therefore, the GCM atmospheres need to be mapped onto a spherical grid with altitude as a vertical coordinate. Because the atmospheres have a different temperature profile in each atmospheric column (with latitude $\alpha$ and longitude $\varphi$), grid points at the same pressure level $P$ do not correspond to the same altitude $z$.

For each latitude and longitude, we calculate the heights associated with the cell centres $\{z\}$ and the cell boundaries $\{\bar{z}\}$:

\begin{align}
\begin{split} \label{eq:interp_altitudes} 
    z_{i}       = {} & \bar{z}_{i} + H \ln ( \bar{P}_i / P_{i} ) \\
    \bar{z}_{i+1} = {} & \bar{z}_{i} + H \ln ( \bar{P}_i / \bar{P}_{i+1} )
\end{split}
\end{align}

\noindent In these equations, symbols with bars are defined on the cell boundaries and those without bars in the cell centres, where it holds that $P_i < \bar{P}_i$. The lower boundary of the bottom cells coincides with the bottom of the atmosphere, such that $\bar{z}_{1}$ = 0 and $\bar{P}_{1}$ = 200 bar. The scale height follows from $H = P_i/\rho_ig$, with $\rho_i$ the local density. The gravity\footnote{In the GCM, a constant gravity $g = GM/R_{\text{0}}^2$ is assumed at all pressures.} is given by $g = GM/(R_{\text{0}}+\bar{z}_i)^2$, with $M$ the mass of the planet and $R_{\text{0}}$ its radius at the bottom of the atmosphere. Once the heights of the cell centres are known, we interpolate the GCM output onto a spherical grid with coordinates ($\alpha$, $\varphi$, $z$), where $\alpha \in [-90^\circ,90^\circ]$ is the latitude, $\varphi \in [-180^\circ,180^\circ)$ is the longitude, and $z$ is the altitude.


As depicted schematically in Figure \ref{fig:day_night_schematic}, the daysides of our models have a substantially larger scale height than the nightsides, owing to the vast temperature difference between both hemispheres. In the strong-drag model, for instance, the atmospheric column that runs from 200 bar to 2 $\mu$bar at the substellar point is a factor $\sim$2.3 taller than at the anti-stellar point. Because the spherical grid goes up to the same altitude everywhere, we are forced to extrapolate the nightsides down to lower pressures, such that the atmosphere is specified at each grid point. To this end, we assume that all cells above the GCM boundary (at 2 $\mu$bar) have temperatures, chemical abundances and wind speeds equal to those in the highest cell \emph{below} the GCM boundary. Densities follow from $\rho_i=P_i/RT_i$, with $R$ the specific gas constant of the atmosphere. We verified that the precise extrapolation assumptions have little effect on the final cross-correlation signals, since the bulk of the absorption occurs at the higher pressures within the GCM boundary (that is, within our selected wavelength windows -- see Section \ref{ss:methods_transit}). 

Another point should be made regarding the wind vectors. In the GCM, the components of the wind vectors are defined as being tangent and orthogonal to isobaric surfaces. However, this is no longer the case once the GCM output has been mapped onto \mbox{($\alpha$, $\varphi$, $z$)} coordinates. Especially in regions where a large temperature gradient occurs (around the terminator), there will be an angle $\gamma$ between the local isobaric surface and the surface of equal altitude. On numerical grounds, however, we expect the impact of this anomaly to be small in relation to the final cross-correlation signals. That is, for all atmospheres, we find that $\gamma \lesssim 20^\circ$ in the cells adjacent to the terminator plane\footnote{For the drag-free model without TiO and VO opacities, we find $\gamma \lesssim 10^\circ$}. The projection factor associated with this angle is $\cos(\gamma) \approx 0.94$, which means that the line-of-sight wind speeds may be off by at most $\sim$$6\%$ in those cells. In practice, the absorption always takes place in multiple cells along a transit chord (some further away from the terminator plane), so the overall error in the Doppler shifts due to winds will be smaller than this value.


\subsection{Monte-Carlo Radiative Transfer with HIRES-MCRT}
\label{ss:methods_mcrt}

In this section, we introduce \textsc{hires-mcrt}, a high-resolution version of the cloudy Monte-Carlo radiative transfer code (\textsc{cmcrt}, \citealt{Lee2017, Lee2019}) that uses high-resolution opacity data instead of binned correlated-$k$ tables. Recently, \citet{Lee2019} performed a series of benchmarking tests, in which the spectra of \textsc{cmcrt} were compared to those of other radiative transfer codes. In Appendix \ref{ap:B}, we benchmark \textsc{hires-mcrt} against CHIMERA (\citealt{Line2013, Line2014}) in the limit of a 1D, uniform atmosphere.

In general, Monte-Carlo radiative transfer (\citealt{Hood2008, Whitney2011,Steinacker2013, Stolker2017, Lee2017, Lee2019, Noebauer2019}) relies on \emph{photon packets}, which perform a random walk throughout a 3D medium of interest. Every time the packet undergoes an interaction, its attributes (e.g. luminosity, direction, polarisation) are updated, until it escapes from the medium. To avoid over-complicating our model and build physical understanding, we do not consider the effects of multiple-scattering in this study, but we leave them for future work. Effectively, we use \textsc{hires-mcrt} as a \emph{randomised transit chord algorithm} (\citealt{Lee2019}), in which the propagation direction of the photon packets does not change after their initialisation. It should be noted that this method is probabilistic and that the spectrum converges to the true solution with an increasing number of photon packets.


As a planet passes in front of its host star, a fraction of the stellar light interacts with the planetary atmosphere, where it falls prey to absorption and scattering. Because the opacity $\kappa$ of the atmospheric constituents is wavelength-dependent, the apparent radius $R_{\text{p}}(\lambda)$ of the planet also varies with wavelength. Measuring $R_{\text{p}}(\lambda)$ at different wavelengths gives rise to the planet's transmission spectrum, which is commonly expressed as a depth $\delta(\lambda) = R^2_\text{p}(\lambda)/R_\star^2$, with $R_\star$ the radius of the host star. To compute $R_{\text{p}}(\lambda)$, \textsc{hires-mcrt} initialises $n$ photon packets for every wavelength in the spectral window of interest ($n = 10^5$ in this work). In case the atmosphere extends from $R_{\text{0}}$ to $R_{\text{0}} + z_{\text{max}}$ (with $z_{\text{max}}$ the maximum altitude), each photon packet is assigned a random impact parameter $b \in (R_{\text{0}}, R_{\text{0}} + z_{\text{max}})$ and impact angle $\Theta \in [0,2\pi)$, in such a way that the atmospheric annulus is uniformly illuminated.

\begin{figure}
\centering
\includegraphics[width=0.48\textwidth]{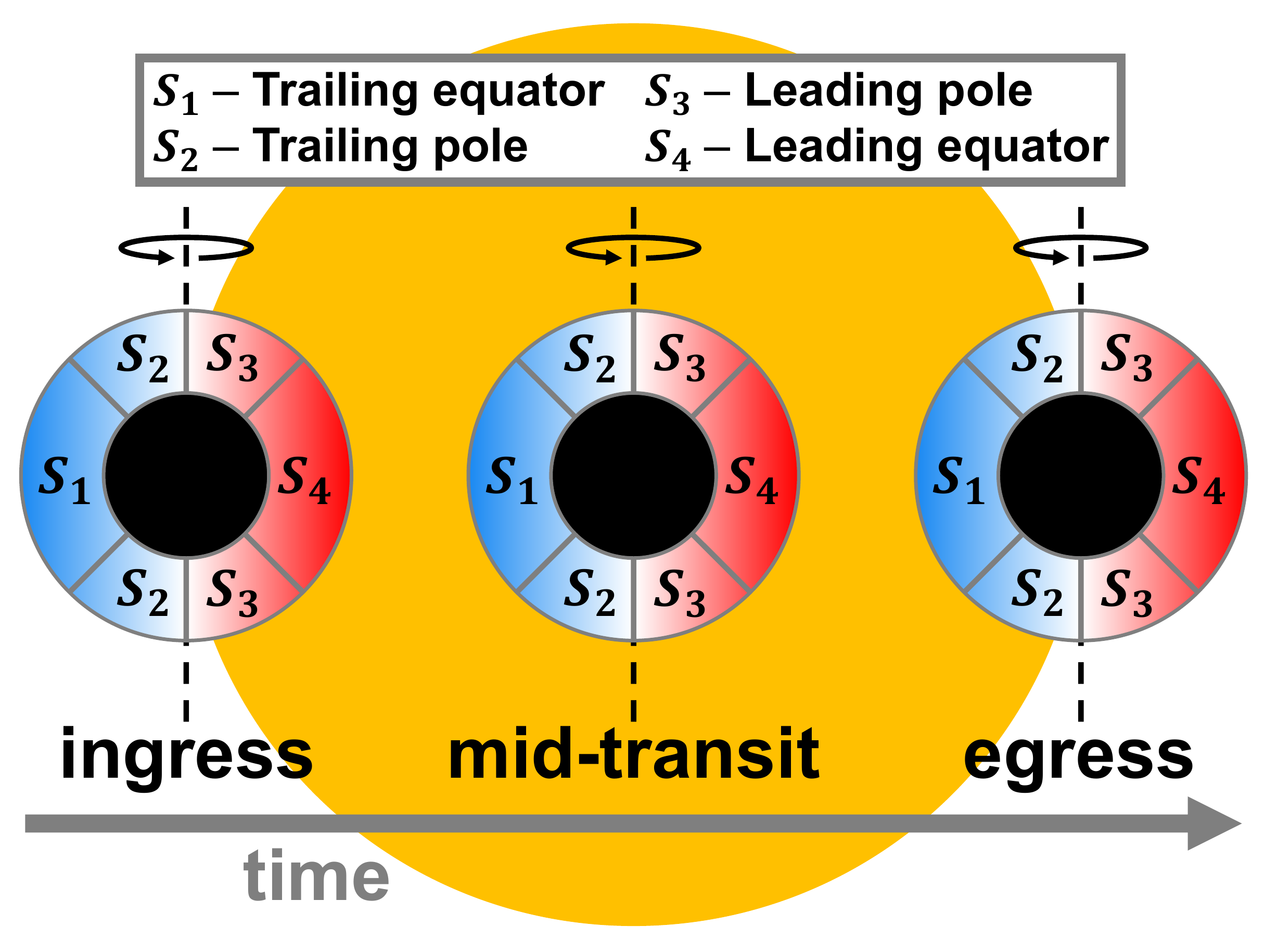}
\caption{Schematic view of a planet at three stages during its transit, with the host star shown in the background. The planet's atmospheric annulus is divided into four limb sectors with equal area: the trailing equator ($\mathcal{S}_1$), the trailing pole ($\mathcal{S}_2$), the leading pole ($\mathcal{S}_3$) and the leading equator ($\mathcal{S}_4$). Because of planetary rotation, opacities along the line of sight are Doppler shifted, resulting in a blueshift on the trailing limb ($\mathcal{S}_1$+$\mathcal{S}_2$) and a redshift on the leading limb ($\mathcal{S}_3$+$\mathcal{S}_4$). Effects of winds are not visualised here.}
\label{fig:limb_sectors_schematic}
\end{figure}

For each photon packet, \textsc{hires-mcrt} evaluates the optical depth $\tau(\lambda, b, \Theta)$ along the corresponding transit chord in a 3D spherical geometry. The optical depth is given by

\begin{equation} \label{eq:tau}
\tau(\lambda, b, \Theta) = \sum_{i=1}^{N_{\text{cells}}} \tilde{\kappa}_i(\lambda, v_{\textsc{los}}) \Delta x_i,
\end{equation}

\noindent where $\tilde{\kappa}_i$ is the opacity (with units cm$^{-1}$) of the $i$-th atmospheric cell and $\Delta x_i$ the distance the photon packet travels through this cell. Note that in high spectral resolution, the effective opacity depends on both the wavelength of the photon packet (in the observer's frame) and the velocity $v_{\textsc{los}}$ of the medium along the line of sight. $v_{\textsc{los}}$ is nonzero due to (i) atmospheric dynamics (i.e. winds), (ii) the rotation of the planet, (iii) its orbital motion and (iv) the system velocity. In this work, however, we assume that all spectra are shifted to the \emph{planetary rest frame}, where the contributions from (iii) and (iv) are zero. Given a particular atmospheric cell, the line-of-sight velocity can thus be written as (e.g., \citealt{Harada2019})


\begin{align}
\begin{split} \label{eq:vlos}
    v_{\textsc{los}} ={}& - u \sin(\varphi+\phi) \\
         & - v \cos(\varphi+\phi)\sin(\alpha) \\
         & + w \cos(\varphi+\phi)\cos(\alpha) \\
         & - \Omega (R_{\text{0}}+z) \sin(\varphi + \phi)\cos(\alpha), \\
\end{split}
\end{align}

\noindent where the first three terms represent the contribution from winds and the last term is due to the rotation of the planet. In Equation \ref{eq:vlos}, $(u,v,w)$ is the cell's wind vector, with $u$ the component in  the local west-to-east (zonal) direction, $v$ the component in the local south-to-north (meridional) direction and $w$ the component in the local nadir-to-zenith (vertical) direction. $\alpha \in [-90^\circ,90^\circ]$, $\varphi \in [-180^\circ,180^\circ)$ and $z \in [0, z_{\text{max}}]$ denote the latitude, longitude and altitude of the cell, respectively. Furthermore, $\phi \in [-180^\circ,180^\circ)$ is the orbital phase angle, which is zero at the primary transit and $\pm180^\circ$ at the secondary eclipse. $\Omega$ denotes the planet's angular frequency (with units rad/s).

\begin{figure*}
\centering
\vspace{-15pt}
\makebox[\textwidth][c]{\hspace{-5pt} \includegraphics[width=1.2\textwidth]{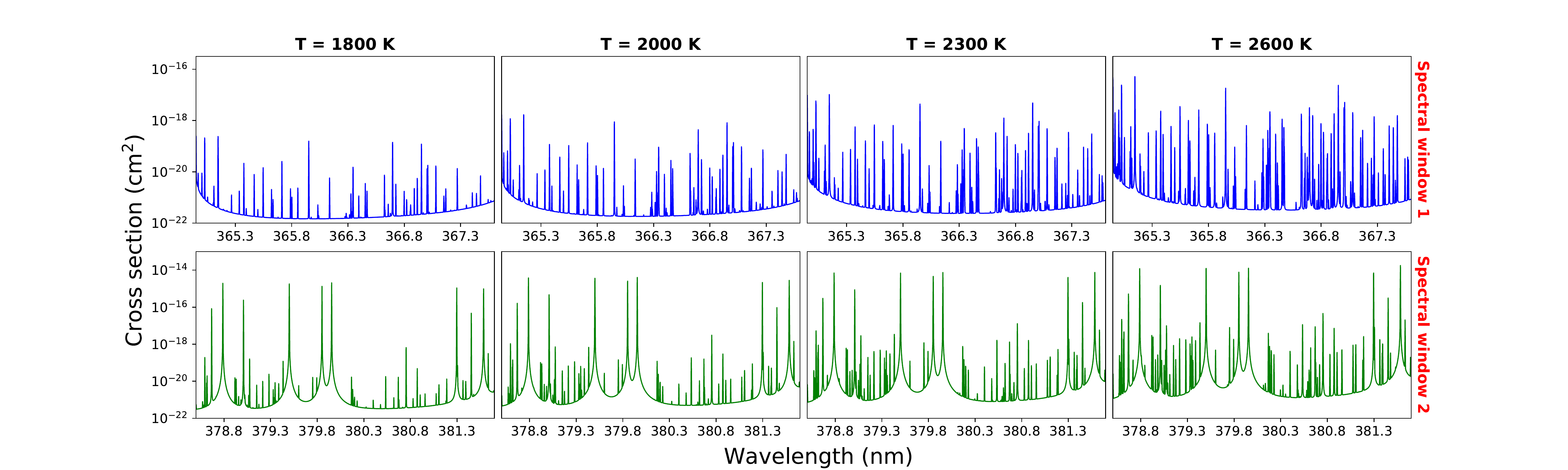}}
\vspace{-10pt}
\caption{The cross-sections of neutral iron (Fe \textsc{i}) at a pressure of 0.3 mbar and four different temperatures (columns). The cross-sections are shown in two wavelength windows (rows), both of which are used to simulate spectra in this work. The top row (window 1) mainly contains weak lines, while the bottom row (window 2) comprises a number of strong lines with broader wings.}
\label{fig:cross_sections}
\end{figure*}

In order for Equation \ref{eq:vlos} to hold, the planet must be tidally locked, such that the sub-stellar point is always given by $\alpha = \varphi = 0$. Also, we assume that the orbit is viewed exactly edge-on (in reality, WASP-76b has an inclination $i=88.0\pm1.6^\circ$). Because Equation \ref{eq:vlos} is key to our modelling efforts, a brief derivation is provided in Appendix \ref{ap:A}, which applies to orbits with arbitrary orientations.

Once the line-of-sight velocity of a cell is known, $\tilde{\kappa}_i$ can be computed by evaluating the opacity function $\kappa_i$ at an effective wavelength $\lambda_\text{eff}$ to account for the associated Doppler shift:

\begin{equation} \label{eq:doppler_shift}
\tilde{\kappa}_i(\lambda, v_{\textsc{los}}) = \kappa_i(\lambda_{\text{eff}}), \ \ \text{with} \ \lambda_{\text{eff}} = \lambda \bigg[ 1-\frac{v_{\textsc{los}}}{c} \bigg].
\end{equation}

\noindent From this equation, it follows that $\lambda_\text{eff} > \lambda$ when the medium moves towards the observer, which seems counter-intuitive. However, from the perspective of the atmospheric cell, $v_{\textsc{los}} < 0$ means that the star is moving away, so the effective wavelength seen by the medium should be redshifted relative to the wavelength seen by the observer.

Once the optical depth is computed along all $n$ transit chords, the transit radius $R_{\text{p}}(\lambda)$ can be found from

\begin{equation} \label{eq:def_transit_radius}
    R^2_{\text{p}}(\lambda) ={} R_{\text{0}}^2 + \Big( (R_{\text{0}} + z_{\text{max}})^2 - R_{\text{0}}^2 \Big) \Big\langle 1 - e^{-\tau} \Big\rangle \bigg\rvert_{\lambda}
\end{equation}

\noindent where the angle brackets denote an average over all photon packets with wavelength $\lambda$. Alternatively, we can write the above equation in terms of areas:

\begin{equation} \label{eq:def_transit_area}
    A_{\text{p}}(\lambda) = A_{\text{0}} + A_{\text{annu}} \big\langle 1 - e^{-\tau} \big\rangle \big\rvert_{\lambda}
\end{equation}

\noindent with $A_{\text{p}}(\lambda)$ the effective area of the planet, $A_{\text{0}}$ the projected area of the planetary interior and $A_{\text{annu}}$ the area of the atmospheric annulus. From Equation \ref{eq:def_transit_area}, it can be readily seen that $A_{\text{p}} = A_{\text{0}}$ when the atmosphere is transparent ($\tau \rightarrow 0$), while $A_{\text{p}} = A_{\text{0}} + A_{\text{annu}}$ when the atmosphere is completely opaque ($\tau \rightarrow \infty$).

\subsection{Decomposing the Atmosphere into Limb Sectors}
\label{ss:methods_limb_sectors}

Since the final spectrum $A_{\text{p}}(\lambda)$ represents an average over the entire -- spatially varying -- limb of the planet, it is hard to identify \emph{how} and to what extent different parts of the atmosphere actually contribute to the overall observable. A way to alleviate this problem is to divide the atmospheric annulus into limb sectors $\mathcal{S}_i$ (see Figure \ref{fig:limb_sectors_schematic}) and to consider the contributions from each of these sectors separately.

One of the advantages of \textsc{hires-mcrt} is that the code can compute transit spectra for any arbitrary subregion of the atmospheric annulus. These spectra can also be found through Equation \ref{eq:def_transit_area}, with the only addition that the average is taken over the transit chords (or, equivalently, photon packets) that cross the sector:

\begin{equation}
    A_{\text{p}, i}(\lambda) = A_{\text{0}} + A_{\text{annu}} \big\langle 1 - e^{-\tau} \big\rangle \big\rvert_{\lambda, \ \mathcal{S}_i}
\end{equation}

\noindent Since we multiply the average extinction with the total area $A_{\text{annu}}$ of the atmospheric annulus, $A_{\text{p},i}(\lambda)$ represents the effective planet area that would be observed if the full limb had the same structure, composition and dynamics as sector $\mathcal{S}_i$.

Suppose we now divide the limb into $N_s$ arbitrary, non-overlapping sectors. In this case, Equation \ref{eq:def_transit_area} can be written as

\begin{equation} \label{eq:spectrum_is_weighted_average}
A_{\text{p}}(\lambda) = \sum_{i=1}^{N_s} \beta_{i} \cdot \bigg( A_{\text{0}} + A_{\text{annu}} \ \big\langle 1 - e^{-\tau} \big\rangle \big\rvert_{\lambda, \ \mathcal{S}_i} \bigg) = \sum_{i=1}^{N_s} \beta_{i} \cdot A_{\text{p},i}(\lambda)
\end{equation}

\noindent with $\beta_{i}$ the fractional area of sector $\mathcal{S}_i$, such that $\sum \beta_{i} = 1$. This result demonstrates that the spectrum of the full limb is a weighted sum of the spectra of the individual sectors. Furthermore, when all limb sectors have the same area (as in Figure \ref{fig:limb_sectors_schematic}), we find that

\begin{equation} \label{eq:average_spectra}
A_{\text{p}}(\lambda) = \Big\langle A_{\text{p},i}(\lambda) \Big\rangle,
\end{equation}

\noindent where the angle brackets denote an average over all limb sectors.

\subsection{Modelling the Transit of WASP-76b}
\label{ss:methods_transit}

We use \textsc{hires-mcrt} to compute self-consistent transmission spectra of the four WASP-76b models (see Section \ref{ss:methods_gcm}) at 37 orbital phases during the transit. Once mapped onto the spherical grid, the atmospheres have a radius $R = R_{\text{0}} + z_{\text{max}} \approx 2.2 R_{\text{Jup}}$, with the 1-bar surface situated between 1.8 and 1.9 $R_{\text{Jup}}$. Additionally, the semi-major axis of the planet's orbit is 0.033 AU. Combined with the radius of the host star, \mbox{$R_\star = 1.73 R_\odot$}, this means that the ingress starts at $ \phi \approx -15.6^\circ$ (first contact) and ends at $\phi \approx -12.1^\circ$ (second contact). Hence, assuming that the system is viewed exactly edge-on, the full transit subtends $\sim$$31^\circ$. Since WASP-76b is tidally locked, this is also the angle by which the atmosphere rotates during the observation. In the ingress and the egress phase, we only illuminate those regions of the atmospheric annulus that are blocking the star.

Transmission spectra are computed in two narrow windows (364.5--367.6 nm and 378.5--381.7 nm) at $R =$ 500,000, which contain 3617 and 4210 wavelength points, respectively. The cross sections of Fe \textsc{i} in these two windows are plotted in Figure \ref{fig:cross_sections}. In the first wavelength window, the opacity is mainly comprised of weak lines that show a considerable temperature-dependence. On the other hand, the second window includes a number of strong lines with broad wings and a fairly constant strength. Because we do not suffer from physical noise in our model, we can afford to limit our study to only a couple of thousand wavelength points for the sake of computational feasibility. In reality, observations require a much larger set of absorption lines to achieve a sufficiently strong signal. This is because the planetary spectrum is typically buried in (stellar) noise.

In \textsc{hires-mcrt}, we account for the cross-sections of Fe \textsc{i} (\citealt{Kurucz1995}), Na (\citealt{Allard2003}), K (\citealt{Allard2016}), VO, TiO, and opacities due to bound-free and free-free transitions associated with H$^-$ (\citealt{John1988}). The absorption cross-sections of the metal oxides, TiO and VO, are from the {\tt EXOPLINES} database  (\citealt{Gharib-Nezhad2021-ZENODO})\footnote{\href{https://zenodo.org/record/4458189\#.YMOoly1h2Al}{Link to EXOPLINES opacity data}} and were generated by \citet{Gharib-Nezhad2021-APJSpaper}, using line lists including \texttt{ExoMol} (\citealt{Tennyson2020}) and \texttt{MoLLIST}  (\citealt{Bernath2020}).

The rationale behind modelling spectra in \emph{two} wavelength windows is the fact that lines of different strength probe the atmosphere at different depths. Hence, spectra in different wavelength windows may be sensitive wind speeds in different regions of the atmosphere. To verify that our simulations are more or less representative of the entire wavelength range spanned by the observations (\citealt{Ehrenreich2020, Kesseli2021}), we computed the CCFs for both windows separately and compared the resulting signals. In spite of some numerical discrepancies, the same trends could be recovered, suggesting that our analysis is at least qualitatively robust. Furthermore, all CCF maps shown in this work (see Sections \ref{s:results_weak_drag} and \ref{s:results_all_gcm}) are obtained by \emph{adding} the maps of both wavelength windows together -- analogous to combining the signals from different spectral orders in real observations (e.g., \citealt{Nugroho2021}).

We note that the opacities of TiO and VO become stronger with increasing wavelength throughout the ESPRESSO and HARPS orders. Hence, if TiO and VO are present in the (observable) atmosphere, the weak iron lines towards the redder end of the spectrum may have reduced amplitudes, which causes the observed CCF signal to be more sensitive to the iron lines in the blue part of the spectrum (simulated in this work). Spectra of an atmosphere \emph{without} TiO and VO do not suffer from this muting effect and may thus contain more weak iron lines at longer wavelengths. We recognise that the models presented in this work do not account for differences in muting or aliasing between models with and without TiO/VO, and this requires more attention in future studies.


\subsection{Computing Cross-Correlation Maps}
\label{ss:methods_ccf}

Once all transmission spectra have been obtained, a 2D cross-correlation map $\text{CCF}(\phi, v)$ can be computed for each atmospheric model. The cross-correlation of a spectrum $\vec{x}(\phi)$ with a template $\vec{T}(v)$ is defined as the inner product of the two:

\begin{equation} \label{eq:def_cross_correlation}
\text{CCF}(\phi, v) = \sum_{j=1}^{N_\lambda} x_j(\phi) \ T_j(v),
\end{equation}

\noindent where the sum is performed over all $N_\lambda$ wavelength points. $\phi$ denotes the orbital phase associated with the spectrum, while $v$ is the velocity (RV) by which the template is shifted. In what follows, we will drop the dependencies on $\phi$ and $v$ for notational convenience. To compute cross-correlation maps of the simulated observations, we use the spectrum of the \emph{static}\footnote{In the \emph{static} models, we set $v_\text{los} = 0$ km/s in all cells.} atmosphere at mid-transit as our template $\vec{T}$. Prior to multiplying the template with the spectra at each of the 37 orbital phases, we subtract its mean such that

\begin{equation} 
\sum_{j=1}^{N_\lambda} T_j = 0.
\end{equation}

\noindent As a result, the template is not sensitive to (constant) offsets in the spectrum, but only to the strength of the spectral features relative to a horizontal baseline. That is,

\begin{equation} \label{eq:mean_zero_template}
\text{CCF} = \sum_{j=1}^{N_\lambda} (x_j + c) \ T_j = \sum_{j=1}^{N_\lambda} x_j \ T_j + c \sum_{j=1}^{N_\lambda} T_j = \sum_{j=1}^{N_\lambda} x_j \ T_j,
\end{equation}

\noindent with $c$ a constant.

Furthermore, because cross-correlation is a linear operation, we can express the CCF map of the full atmospheric limb as the average of the CCF maps associated with the individual limb sectors $\mathcal{S}_i$. Since the transit depth is proportional to the effective area of the planet, we can substitute Equation \ref{eq:spectrum_is_weighted_average} into Equation \ref{eq:def_cross_correlation} to obtain

\begin{equation}
\text{CCF} = \sum_{j=1}^{N_\lambda} \bigg( \sum_{i=1}^{N_s} \beta_i x_{i,j} \bigg) \ T_j,
\end{equation}

\noindent with $x_{i,j}$ the spectral value of sector $\mathcal{S}_i$ at wavelength $j$. Rearranging gives

\begin{equation} \label{eq:ccf_is_weighted_average}
\text{CCF} = \sum_{i=1}^{N_s} \beta_i \cdot \bigg( \sum_{j=1}^{N_\lambda} x_{i,j} \ T_j \bigg) = \sum_{i=1}^{N_s} \beta_i \cdot \text{CCF}_i,
\end{equation}

\noindent with CCF$_i$ the cross-correlation that is obtained when multiplying the spectrum of sector $\mathcal{S}_i$ with the template. By analogy with Equation \ref{eq:average_spectra}, it holds that 

\begin{equation} \label{eq:ccf_is_average}
\text{CCF} = \Big\langle \text{CCF}_i \Big\rangle,
\end{equation}

\noindent when all sectors have the same area, as is the case in this work. Equation \ref{eq:ccf_is_average} is an important result, because it allows us to decompose the CCF map of the atmosphere into multiple contributions that are easier to interpret than the signal of the full limb. The total CCF is simply proportional to the sum of these contributions.

We remind the reader that Equation  \ref{eq:def_cross_correlation} is not a unique definition of the CCF. Another frequently used expression is the Pearson correlation coefficient, where the inner product from Equation \ref{eq:def_cross_correlation} is normalised by the square root of the product of the spectrum and template variances (see Appendix B in \citealt{Hoeijmakers2019} for further details). In this work, we do not use the Pearson correlation coefficient because it does not satisfy Equation \ref{eq:ccf_is_weighted_average}. This is because different sectors may have spectra with a different mean value, such that the associated normalisation factor changes for every sector, and may not be equal to that of the total CCF. Also, we want to keep track of variations in the signal strength across the transit, and this information would be lost if we used the Pearson correlation coefficient.

\begin{figure*}
\centering

\vspace{-20pt}

\makebox[\textwidth][c]{\hspace{-30pt} \includegraphics[width=1.2\textwidth]{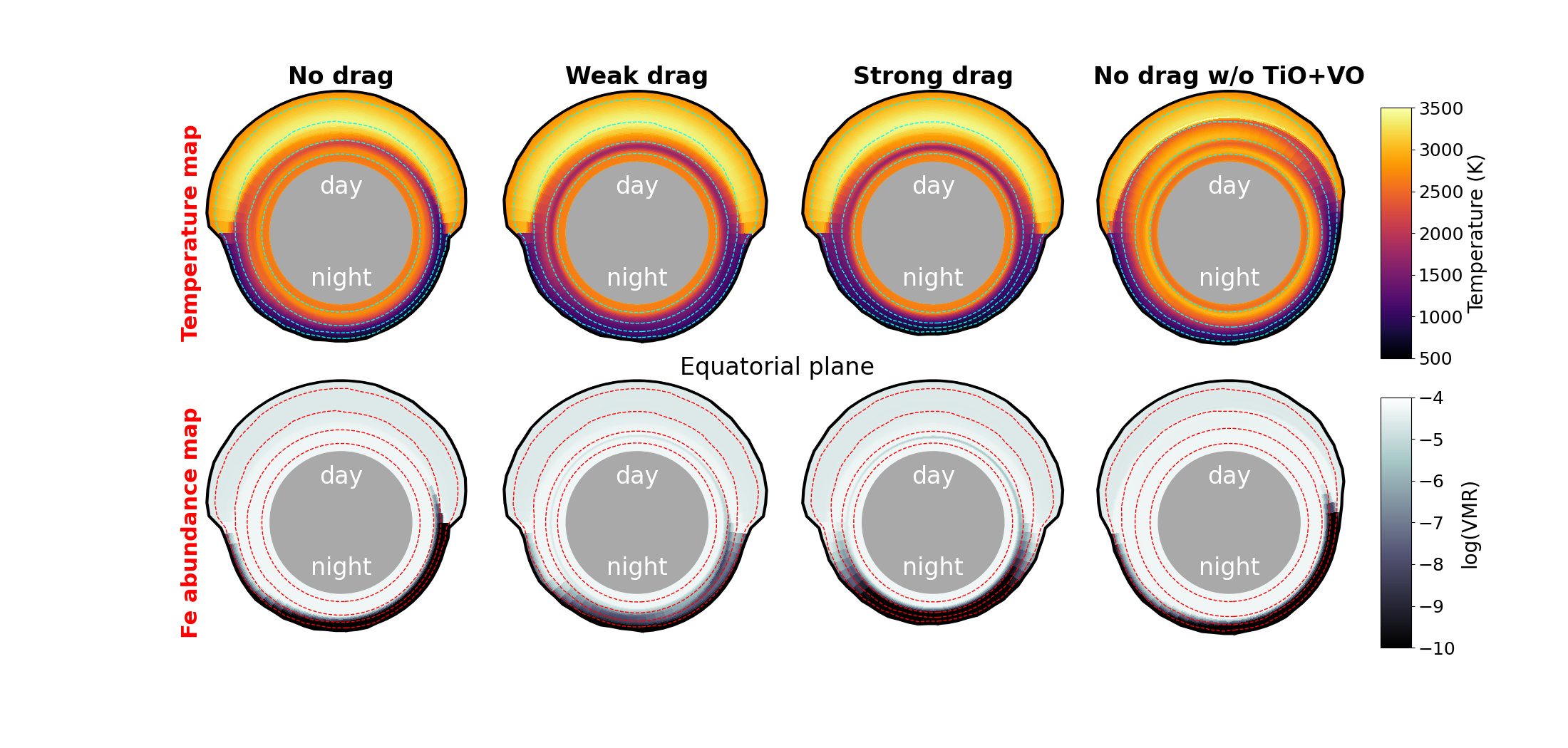}}

\vspace{-41pt}

\caption{\textbf{Top row:} Temperatures in the equatorial plane of the of the four GCM atmospheres listed in Table \ref{tab:gcm_models}. The maps are plotted as a function of spatial coordinate, with the dashed lines denoting isobars of $10^1$, $10^{-1}$, $10^{-3}$ and $10^{-5}$ bar, respectively. The solid black contour marks the GCM boundary at 2 $\mu$bar. \textbf{Bottom row:} Iron abundances in the equatorial plane, expressed in terms of volume mixing ratio (VMR). Note that the size of the atmosphere with respect to the rest of the planet has been enhanced for visualisation purposes.}
\label{fig:gcm_equator_temp_vmr}
\end{figure*}

\subsection{Computing $K_\text{p}$--$V_\text{sys}$ Maps}
\label{ss:methods_kpvsys}

To produce the CCF maps discussed in the previous paragraph, we multiply the planet spectra $\vec{x}(\phi)$ at different orbital phases with a template $\vec{T}(v)$ shifted by many different velocities. For a planet with a circular orbit, there exists a simple relationship between its radial velocity $v$ and its orbital phase angle $\phi \in [-180^\circ,180^\circ)$:

\begin{equation}
v = V_{\text{sys}} + K_{\text{p}} \sin(\phi),
\end{equation}

\noindent with $V_{\text{sys}}$ the system velocity and $K_{\text{p}}$ the velocity semi-amplitude. Through this equation, it is also possible to express the cross-correlation as a function of $K_{\text{p}}$ and $V_{\text{sys}}$. That is, CCF = CCF$(K_{\text{p}}, V_{\text{sys}}, \phi)$. In practice, the cross-correlation is always a discrete function, so it can be represented by a cube of numbers CCF$_{j,k,l}$, where $j$, $k$ and $l$ denote the $K_{\text{p}}$, $V_{\text{sys}}$ and $\phi$ axes, respectively.

In order to combine the spectra from all orbital phases, and obtain a single signal-to-noise ratio (SNR) for the \emph{entire} observation, one typically co-adds the CCF values along the phase axis:

\begin{equation} \label{eq:def_kpvsys}
\text{SNR}(K_{\text{p}}, V_{\text{sys}}) \equiv \sum_{l=1}^{N_\phi} a \ \big( \text{CCF}_{j,k,l} - b \big),
\end{equation}

\noindent with $a$ a scaling factor and $b$ a baseline (\citealt{Hoeijmakers2015, Cabot2020}). In the absence of atmospheric dynamics and planetary rotation, the SNR must acquire a maximum at the true \mbox{($K_{\text{p}}$, $V_{\text{sys}}$)} values of the planet, when all CCF peaks in the data cube align. The value of this maximum is the SNR associated with the detection of the species whose absorption lines were included in the template. It should be stressed that the SNR is a \emph{time-averaged} quantity, because we co-add the CCF values from different orbital phases.

As with the CCF, we can also express the SNR in terms of limb sectors. Plugging Equation \ref{eq:ccf_is_weighted_average} into Equation \ref{eq:def_kpvsys} yields

\begin{align} \label{eq:kpvsys_mean}
\begin{split}
    \text{SNR} ={}& \sum_{l=1}^{N_\phi} a \ \Bigg[ \Bigg( \sum_{i=1}^{N_s} \beta_i \cdot \text{CCF}_{i,j,k,l} \Bigg) - b \Bigg] \\
         ={} & \sum_{l=1}^{N_\phi} \Bigg[ a \ \sum_{i=1}^{N_s}  \beta_i \ \big(\text{CCF}_{i,j,k,l} - b \big) \Bigg] \\
         ={} & \sum_{i=1}^{N_s} \Bigg[ \beta_i \sum_{l=1}^{N_\phi} a \ \big(\text{CCF}_{i,j,k,l} - b \big) \Bigg] = \sum_{i=1}^{N_s} \beta_i \cdot \text{SNR}_i \\
\end{split}
\end{align}

\noindent where we dropped the dependencies on $K_{\text{p}}$ and $V_{\text{sys}}$. Furthermore, $\text{SNR}_i$ is the signal-to-noise ratio associated with sector $\mathcal{S}_i$. Note that equation \ref{eq:kpvsys_mean} only holds when the same values of $a$ and $b$ are used for every sector. Furthermore, when all limb sectors have the same area, we find that

\begin{equation} 
\text{SNR} = \Big\langle \text{SNR}_i \Big\rangle.
\end{equation}

\noindent In this work, we calculate the $K_{\text{p}}$--$V_{\text{sys}}$ maps based on the 29 \emph{in-transit} spectra, so we ignore the ingress and egress phase. This prevents us from having to deal with limb sectors that are only partially illuminated by the star, which complicates the treatment.


\section{GCM structures and Transmission Spectra}
\label{s:spec_and_struc} 

\subsection{Four Atmospheric Models of WASP-76b}

Figure \ref{fig:gcm_equator_temp_vmr} shows the temperatures and iron abundances in the equatorial plane of the WASP-76b models obtained from the GCM. As mentioned previously, the dayside is substantially more extended than the nightside, due to the large temperature contrast between both hemispheres. It can be seen that the nightside of the drag-free atmospheres is slightly warmer (especially between 10 and 0.1 bar) compared to the models with drag, owing to more efficient heat redistribution. Furthermore, the dayside of the first three models features a strong thermal inversion between 0.1 and $10^{-3}$ bar, where the temperature increases with altitude. The plots also reveal a hotspot shift, a displacement of the hottest region away from the sub-stellar point. In the drag-free scenarios, the atmospheric layers between 0.1 and $10^{-3}$ bar are hotter to the east of the sub-stellar point than to the west. However, as the drag timescale becomes shorter, the hotspot shift diminishes -- a result that is in agreement with the literature (\citealt{Showman2011}, \citealt{Parmentier2017}, \citealt{Tan2019}).

Because the drag-free model without TiO and VO opacities lacks important optical absorbers, its temperature structure is qualitatively different from that of the other models. In particular, its hotspot offset extends to much lower pressures $\sim$$10^{-4}$ bar, resulting in a larger asymmetry between the trailing and leading limb of the planet. As shown in the right panels of Figure \ref{fig:gcm_equator_temp_vmr}, the day-night transition to smaller scale heights appears to be more gradual on the leading limb than on the trailing limb. The atmosphere also has a steep thermal inversion on its dayside, but it occurs higher up in the atmosphere compared to the other models, at pressures below $10^{-3}$ bar. This inversion is likely due to sodium absorption in the upper atmosphere.

As far as chemistry is concerned, the bottom panels in Figure \ref{fig:gcm_equator_temp_vmr} illustrate that iron is more or less uniformly distributed over the dayside of the WASP-76b models. On the cooler nightside, however, the upper regions of the atmosphere are (almost) fully depleted of iron as a result of condensation. As discussed in Section \ref{ss:methods_gcm}, the GCM accounts for condensation through rainout.

\begin{figure*}
\centering

\vspace{-20pt}

\makebox[\textwidth][c]{\hspace{-30pt} \includegraphics[width=1.2\textwidth]{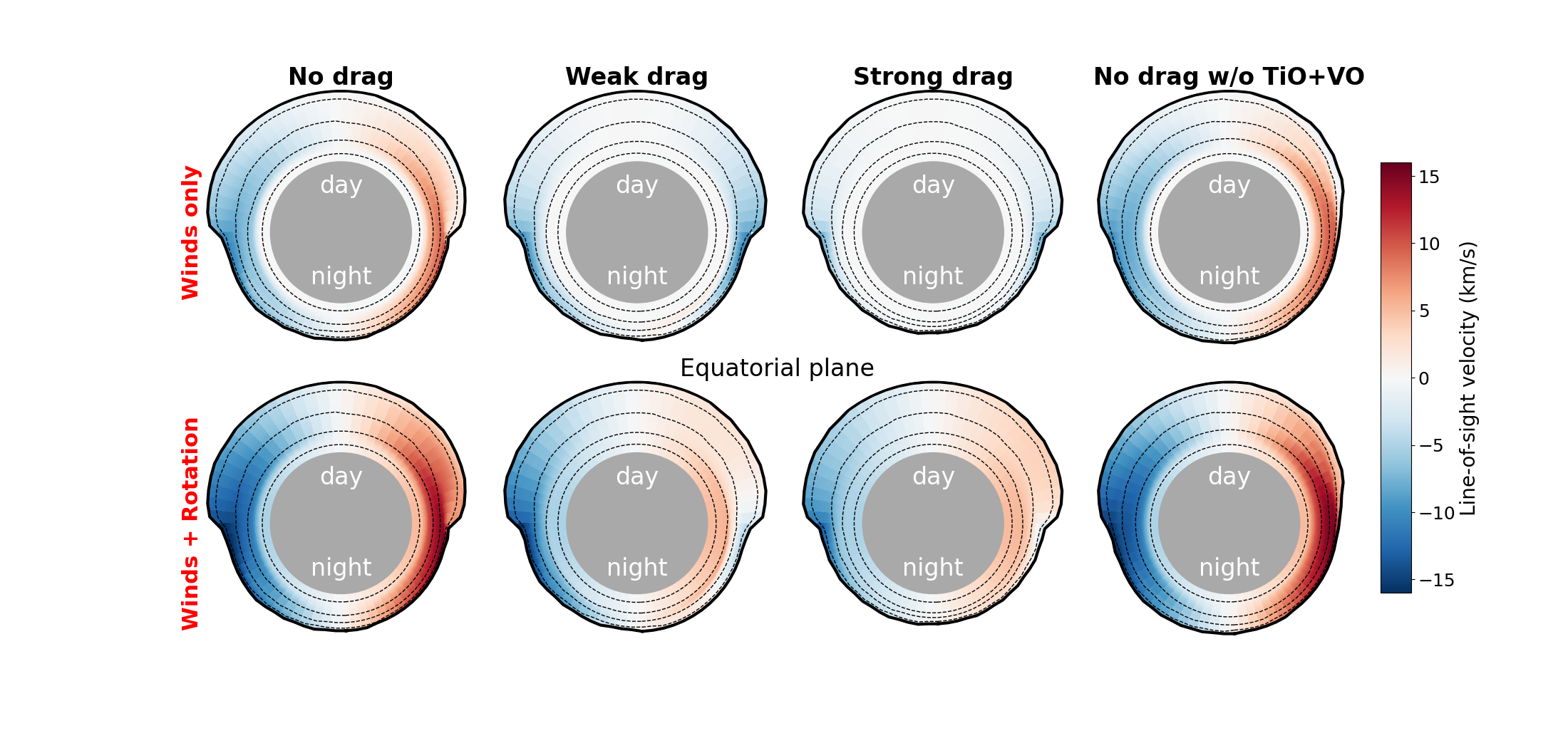}}

\vspace{-41pt}

\caption{\textbf{Top row:} Line-of-sight velocities due to winds in the equatorial plane of the of the four GCM atmospheres listed in Table \ref{tab:gcm_models}. A redshift (blueshift) indicates that the winds are blowing away from (towards) the observer. The maps are plotted as a function of spatial coordinate, with the dashed lines denoting isobars of $10^1$, $10^{-1}$, $10^{-3}$ and $10^{-5}$ bar, respectively. The solid black contour marks the GCM boundary at 2 $\mu$bar. \textbf{Bottom row:} The same plots, but with the contribution from planetary rotation added.}
\label{fig:gcm_equator_vlos}
\end{figure*}

\begin{figure*}
\centering

\vspace{-20pt}

\makebox[\textwidth][c]{\hspace{-30pt} \includegraphics[width=1.2\textwidth]{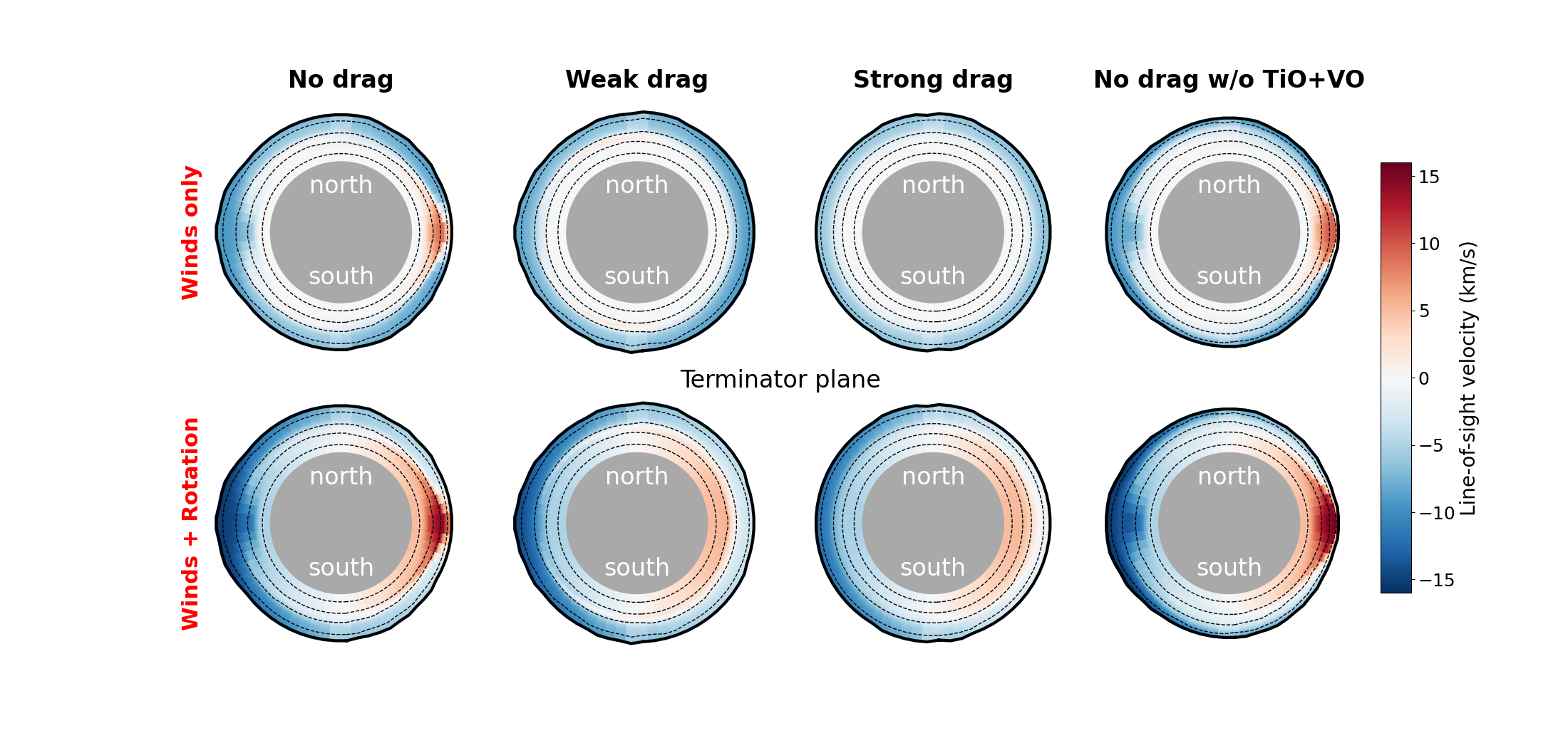}}

\vspace{-41pt}

\caption{\textbf{Top row:} Line-of-sight velocities due to winds in the terminator plane of the of the four GCM atmospheres listed in Table \ref{tab:gcm_models}. A redshift (blueshift) indicates that the winds are blowing away from (towards) the observer. Dashed lines denote isobars, similar to Figures \ref{fig:gcm_equator_temp_vmr} and \ref{fig:gcm_equator_vlos}. \textbf{Bottom row:} The same plots, but with the contribution from planetary rotation added. Again, note that the overall extent of the atmospheres has been exaggerated for visualisation purposes.}
\label{fig:gcm_terminator_vlos}
\end{figure*}

Figures \ref{fig:gcm_equator_vlos} and \ref{fig:gcm_terminator_vlos} show the line-of-sight velocities of the atmospheres in the equatorial plane and the terminator plane, respectively. We distinguish between the line-of-sight velocities due to winds only (top rows) and those due to the combined effect of winds and rotation (bottom rows). A number of things can be pointed out. Firstly, the models without drag develop an eastward equatorial jet. As a consequence, the associated line-of-sight velocities are negative on the trailing limb and positive on the leading limb (Figure \ref{fig:gcm_equator_vlos}). Away from the equatorial plane, however, the terminator is predominantly blueshifted due to the prevalence of day-to-night winds (Figure \ref{fig:gcm_terminator_vlos}). When drag forces are introduced, the \emph{full} terminator appears blueshifted, because jet formation around the equator is inhibited. Also, the shorter the drag timescale becomes, the lower resulting the wind speeds.

\begin{figure*}
\vspace{-10pt}
\centering
\makebox[\textwidth][c]{\includegraphics[width=1.1\textwidth]{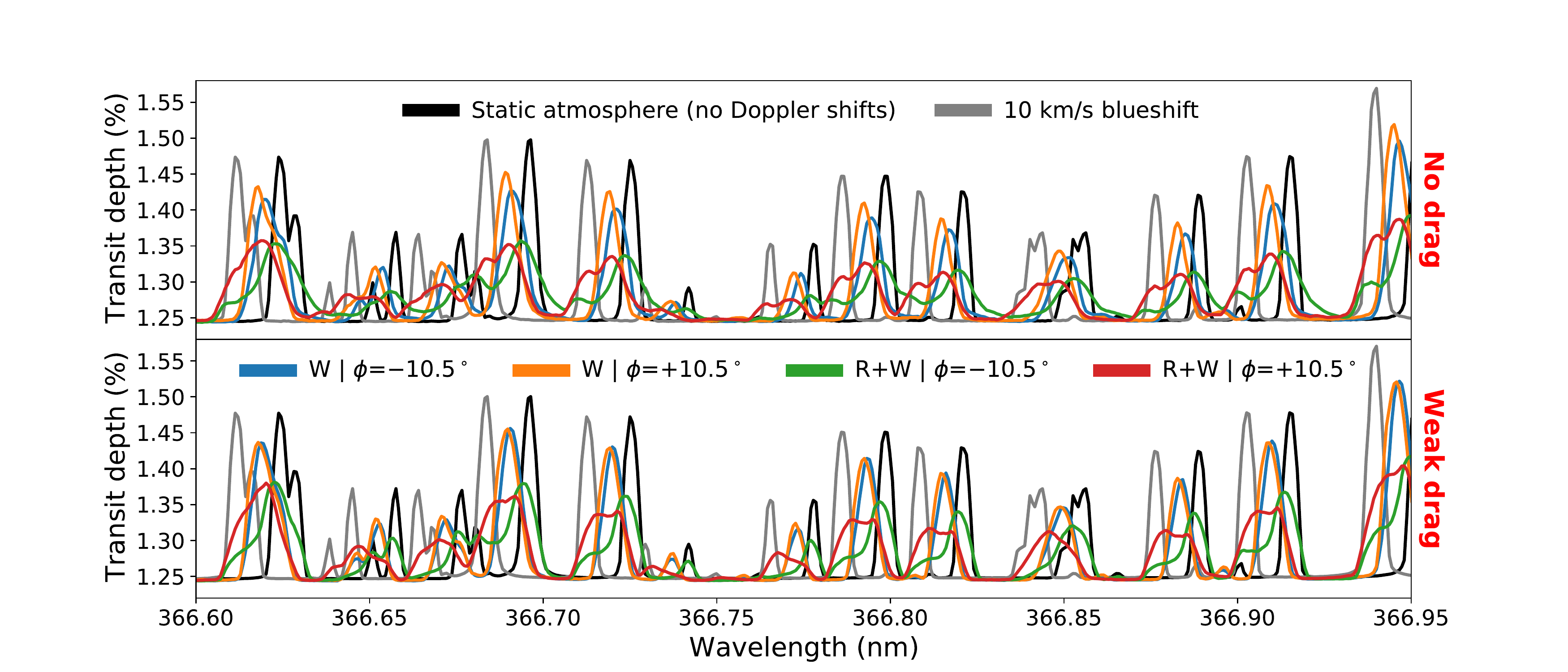}}
\vspace{-15pt}
\caption{Transmission spectra of the drag-free (top panel) and the weak-drag (bottom panel) atmospheres, at phase angles $-10.5^\circ$ and $+10.5^\circ$. For both orbital phases, two spectra are plotted: one (\texttt{W}) that only includes Doppler shifts due to winds, and one (\texttt{R+W}) that includes both the effects of winds and rotation. The spectrum of the static atmosphere, with $v_{\textsc{los}} = 0$ km/s in all cells, is plotted in the background. As a reference, the figure also shows the static spectrum offset by $-$10 km/s.}
\label{fig:wasp76b_spectra}
\end{figure*}

When planetary rotation enters the equation, the atmosphere's trailing limb (which rotates towards the observer) picks up an additional blueshift, while the leading limb (which rotates away from the observer) acquires a redshift. It should be noted that the contribution from rotation to the line-of-sight velocities is greatest in the equatorial regions, which lie furthest away from the rotation axis. Around the equator, the rotational velocity of WASP-76b is $\pm5.3$ km/s. At the poles, the contribution is zero.

To provide further insight into the structure of the atmospheres, we present four more figures like Figure \ref{fig:gcm_terminator_vlos} in Appendix \ref{ap:C}, where we show the interplay between chemistry, temperature and wind speed in the limb plane of the planet (perpendicular to the line of sight). Additionally, we show how the limb plane changes between ingress, mid-transit and egress, to illustrate the effects of planetary rotation.

\begin{figure*}
\centering

\makebox[\textwidth][c]{\hspace{-10pt} \includegraphics[width=0.88\textwidth]{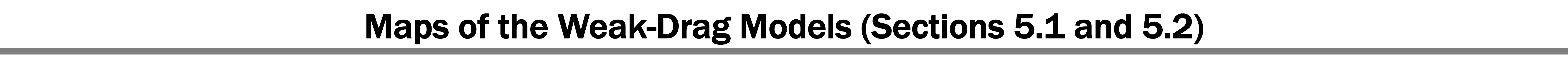}}

\vspace{-40pt}

\makebox[\textwidth][c]{\includegraphics[width=1.2\textwidth]{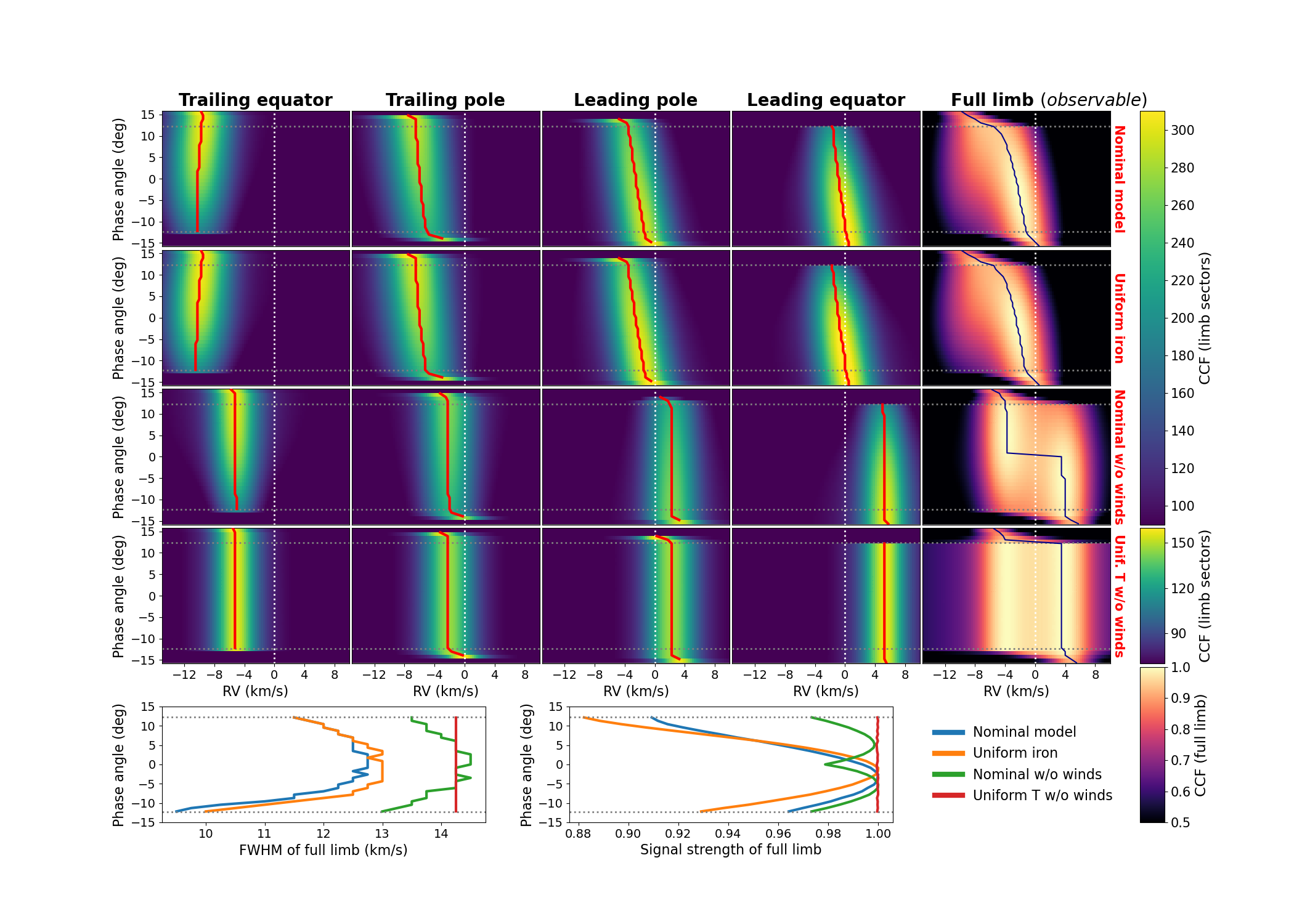}}

\vspace{-30pt}

\caption{\textbf{Top four rows}: CCF maps of four atmospheres, all based on the weak-drag model from the GCM. The ``nominal'' model (top row) is the weak-drag model with Doppler shifts due to winds and rotation. The other atmospheres were obtained by making simple, ad hoc changes to the structure of the weak-drag model (see text). The first four columns pertain to the different limb sectors of the atmospheric annulus (see Figure \ref{fig:limb_sectors_schematic}), while the right column shows the CCF maps of the full limb, normalised such that their maximum value is unity. Red and dark blue curves indicate the maxima of each map as a function of orbital phase. The horizontal dotted lines mark the end of the ingress and the start of the egress, respectively. \textbf{Bottom-left panel}: FWHM of the CCF maps of the full limb as a function of orbital phase. \textbf{Bottom-right panel}: Maxima of the CCF maps of the full limb as a function of orbital phase.}
\label{fig:ccf_panels1}
\end{figure*}

\subsection{Transmission Spectra}

Figure \ref{fig:wasp76b_spectra} shows a number spectra computed with \textsc{hires-mcrt} (for the drag-free and weak-drag atmospheres), in a small section of the first wavelength window. We distinguish between spectra that only include Doppler shifts due to winds (first three terms in Equation \ref{eq:vlos}), and spectra including the effects of both winds and planetary rotation (all terms in Equation \ref{eq:vlos}). A number of things can be pointed out. Firstly, all spectra are blueshifted by several km/s compared to the spectrum of the static atmosphere (with $v_{\textsc{los}} = 0$ km/s everywhere). Also, the absorption lines in the wind-only spectra are slightly broadened due to a dispersion in the (line-of-sight) wind speeds across the atmospheric limb. Planetary rotation further broadens the absorption features, as opposite redshift and blueshift contributions tend to pull the lines apart. Finally, Figure \ref{fig:wasp76b_spectra} demonstrates that the spectra at $\phi = -10.5^\circ$ are different from those at $\phi = +10.5^\circ$. The shapes, positions and depths of the lines change during the transit, and this is a direct effect of the 3D geometry of the observation -- different parts of the atmosphere are probed at different orbital phases, resulting in a time-dependent signal.


\section{CCF Maps of the Weak-Drag Model}
\label{s:results_weak_drag}

Since \citet{Ehrenreich2020} suggested that the atmosphere of \mbox{WASP-76b} only features day-to-night winds (and no equatorial jet), we start by examining the CCF maps of our weak-drag model, with $\tau_\text{drag} = 10^5$ s. In this section, we aim to interpret the structure of the maps and characterise the behaviour of the signals arising from different limb sectors.

\subsection{Weak Drag: the Nominal Model}
\label{ss:results_weak_drag_nominal}

The top row of Figure \ref{fig:ccf_panels1} displays the CCF maps obtained for the weak-drag model of WASP-76b, where we account for Doppler shifts due to winds and planetary rotation. The top-right panel shows the signal corresponding to the full limb (observable in reality). Its maximum shifts from $-$1 km/s at the end of ingress to $-$5.5 km/s at the start of egress. In addition to the \emph{location} of the CCF peak, the full width at half maximum (FWHM) and the normalised peak height (signal strength) also change over time, as illustrated in the bottom panels of Figure \ref{fig:ccf_panels1}. It should be noted that these quantities were computed after the subtraction of single baseline value from the entire CCF map.

It is no surprise that the planet's cross-correlation map is time-dependent. Because the planet rotates by $\sim$30 degrees during its transit, the observation really probes different parts of the atmosphere at different orbital phases -- especially around the equator. Furthermore, the polar regions, which are situated close to the rotation axis, are probed under different angles over time. This means we can expect the observation to be sensitive to spatial variations in the planet's temperature and abundance maps, as well as to the magnitude and orientation of the wind profile.

\begin{figure*}
\centering
\vspace{-20pt}
\makebox[\textwidth][c]{\hspace{-30pt} \includegraphics[width=1.26\textwidth]{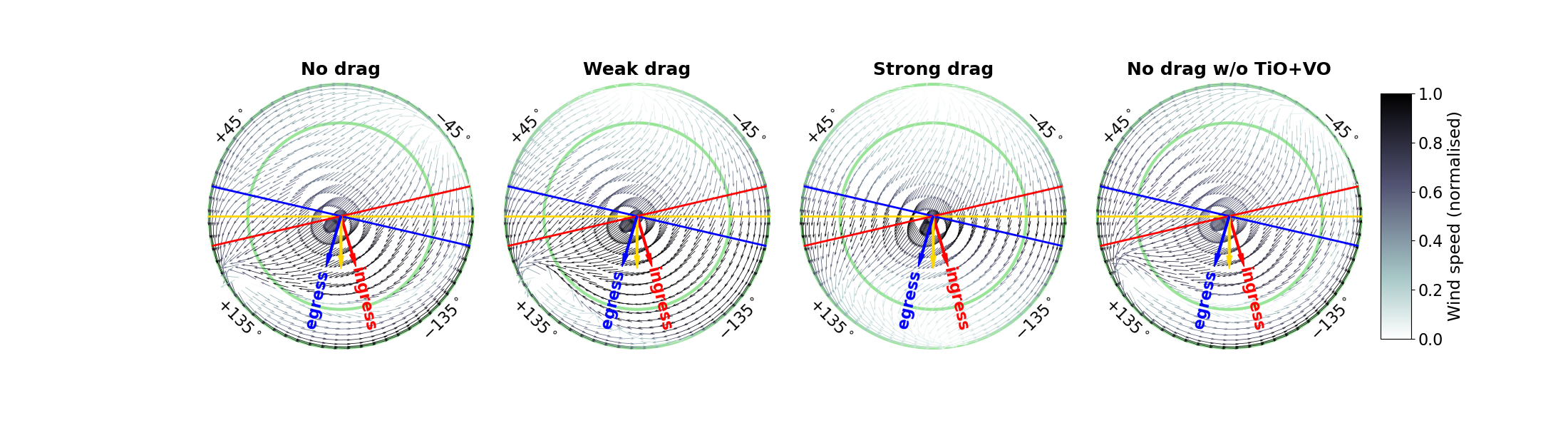}}
\vspace{-31pt}
\caption{Wind profiles (at $P=$ 0.15 mbar) of the four GCM models from Table \ref{tab:gcm_models}, projected onto their northern hemisphere. We are looking down onto the planet from a point above its north pole. Each of the profiles is normalised to its own maximum. The darker the colour of the vectors, the higher the associated wind speeds. The green circles denote 0 and 45 degrees latitude. The solid lines represent the planet's limb plane at ingress (red), mid-transit (yellow) and egress (blue), respectively. The plots demonstrate that the angle between the polar wind vectors and the line of sight (coloured arrow perpendicular to the limb plane) becomes smaller as the planet rotates from ingress to egress.}
\label{fig:gcm_windvectors}
\end{figure*}

\begin{figure}
\centering
\vspace{-28pt}

\makebox[0.5\textwidth][l]{\hspace{-7pt} \includegraphics[width=0.51\textwidth]{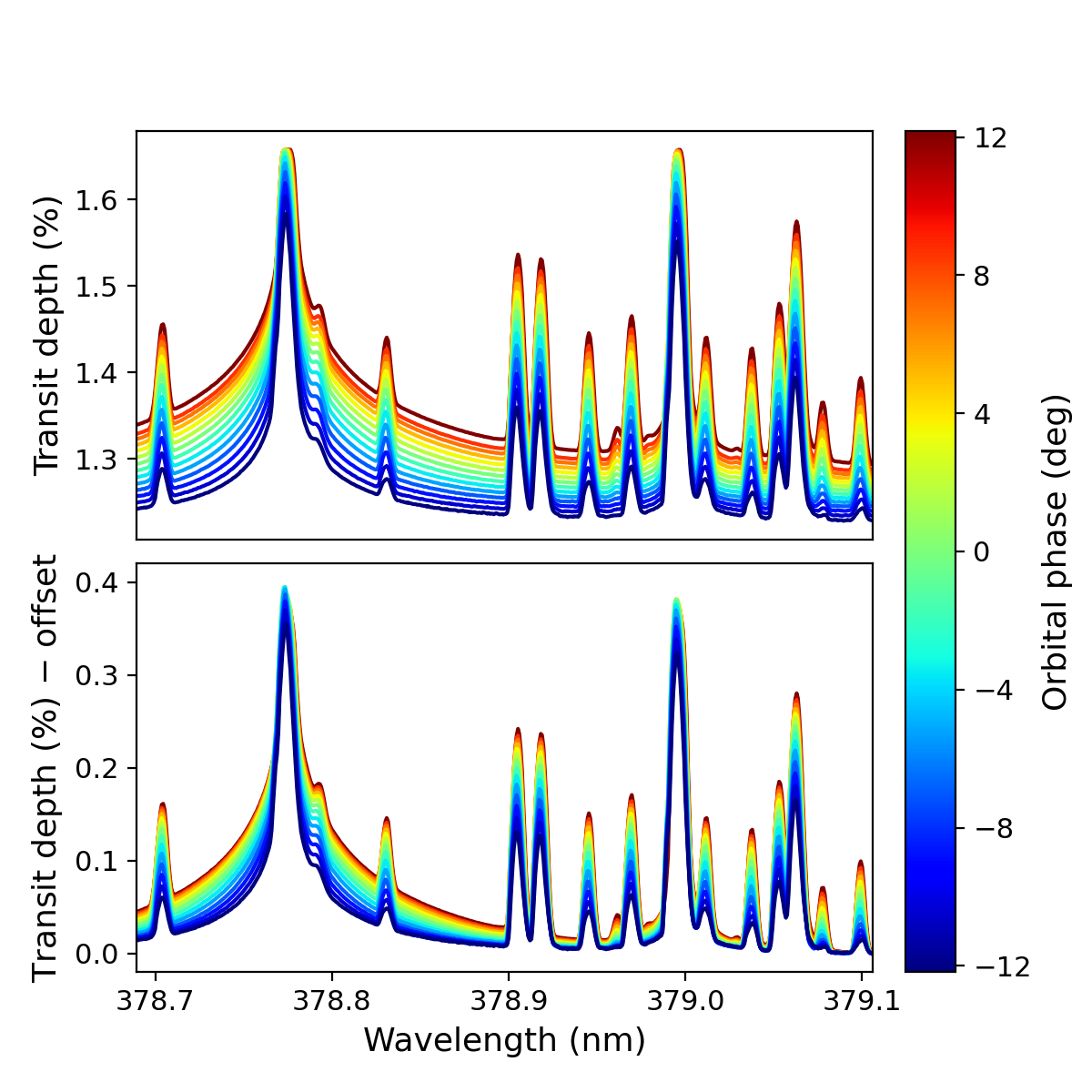}}

\vspace{-10pt}
\caption{\textbf{Top:} Trailing-equator spectra of the weak-drag atmosphere (nominal model in Figure \ref{fig:ccf_panels1}) in a small section of the simulated wavelength range. The colour scale indicates different orbital phase angles. \textbf{Bottom:} The same spectra, but with their baselines subtracted.}
\label{fig:trailing_limb_spec}
\end{figure}

To better understand the structure of the CCF map associated with the full limb, we consider the CCF maps arising from each of the four limb sectors depicted in Figure \ref{fig:limb_sectors_schematic} -- the trailing equator, the trailing pole, the leading pole and the leading equator. As demonstrated in Section \ref{ss:methods_ccf}, the CCF map of the full limb is simply the average of the (less complicated) CCF maps of the individual sectors. Hence, Figure \ref{fig:ccf_panels1} allows us to interpret the different contributions to the overall signal separately, and link them to specific regions in the atmosphere.

The first thing to note is that each of the limb sectors has a different RV offset from the planetary rest frame. Moving from the left to the right panel in the first row of Figure \ref{fig:ccf_panels1}, one can see that the RV of the sectors becomes less negative. This trend is a direct effect of planetary rotation, which induces a blueshift on the trailing limb and a redshift on the leading limb. Also, the further away a sector from the rotation axis, the larger the magnitude of the Doppler shift associated with rotation. Besides rotation, day-to-night winds also contribute to the RV offset, which is why all sectors are blueshifted. 

The \emph{slopes} in the RV signals of the sectors are due to the planet's wind profile. On the trailing limb, the signal becomes marginally more redshifted with time, while the other sectors show an increasing blueshift with time. Around the poles, where roughly the same regions are probed during the transit, we are likely witnessing a projection effect. This is illustrated in Figure \ref{fig:gcm_windvectors}. As the planet rotates between ingress and egress, the angle between the polar wind vectors and the line of sight becomes smaller, resulting in a more negative line-of-sight velocity with orbital phase. Near the equator, atmospheric regions move into and out of view more quickly, so the observation could actually probe different wind speeds there.

Besides the slopes and offsets of the RV shifts, the \emph{signal strengths} also change in every sector -- most noticeably around the equator. On the trailing equator, for instance, the CCF peak height increases during the transit. Figure \ref{fig:day_night_schematic} illustrates why such behaviour is inherent to the atmospheres of ultra-hot Jupiters. Just after the ingress, when stellar light first crosses the trailing equator of the planet, the observation probes a relatively large part of the nightside (top left panel in Figure \ref{fig:day_night_schematic}). Yet, as the transit progresses, a larger part of the dayside rotates into view and the absorption takes place in progressively hotter regions. This has two consequences for the spectrum of the trailing equator, as demonstrated in Figure \ref{fig:trailing_limb_spec}: (i) the baseline moves up as the dayside is more puffy, and (ii) the absorption lines become stronger relative to the baseline. The reason for these stronger absorption lines is again twofold. Firstly, the weak iron lines become stronger with increasing temperature (see Figure \ref{fig:cross_sections}). Secondly, the scale height is larger in the hotter parts of the atmosphere, which results in bigger transit-depth variations with wavelength. In addition, the strong iron lines (one of which is shown in Figure \ref{fig:trailing_limb_spec}) acquire broader wings. Because the CCF is sensitive to line strength, its value must increase during the transit. The exact opposite occurs on the leading equator, which initially comprises a substantial part of the hot dayside. However, over the course of time, the nightside rotates into view and the absorption takes place in cooler regions with a smaller scale height. Similar effects occur around the poles, but the variations are less extreme, as the sectors are closer to the planet's rotation axis.

\begin{figure*}
\centering
\vspace{-10pt}

\makebox[\textwidth][c]{\hspace{-40pt} \includegraphics[width=1.2\textwidth]{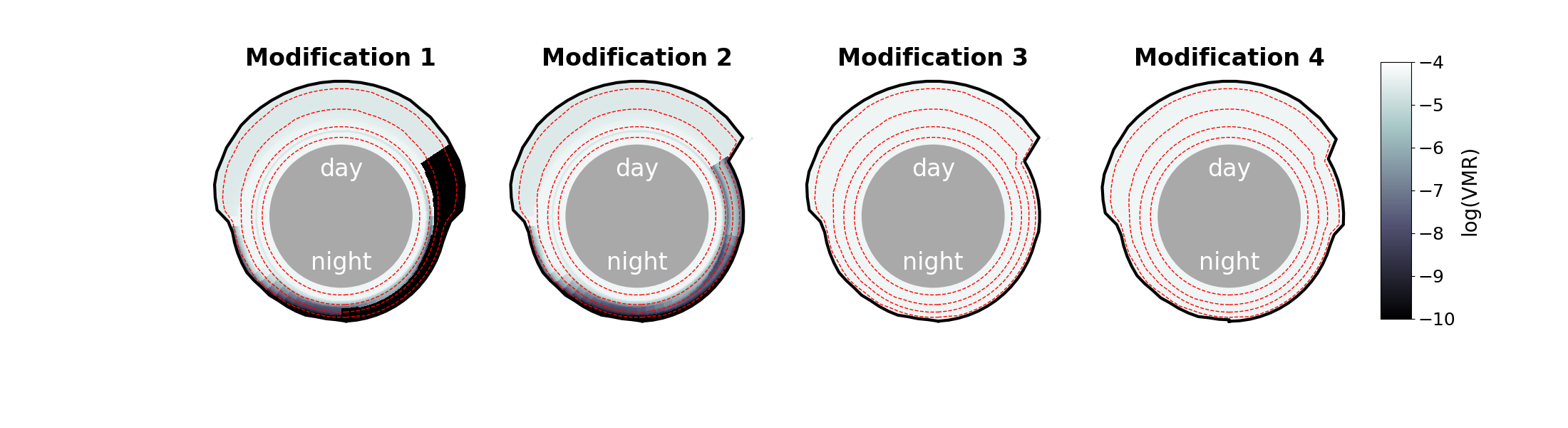}}
\vspace{-45pt}
\caption{Iron abundances in the equatorial plane of the modified weak-drag models discussed in Section \ref{ss:results_weak_drag_iron_removed}. In modification 1, all gaseous iron was removed from cells with longitudes $-180^\circ < \varphi < -58^\circ$ and pressures $P < 0.01$ bar. In modifications 2 and 3, the nightside temperature structure was extended up to $-58^\circ$ longitude, resulting in a smaller scale height on the leading limb. In modification 4, the leading limb has a constant temperature of 1800 K.}
\label{fig:modified_vmr}
\end{figure*}

\subsection{Weak Drag: Variations on the Nominal Model}
\label{ss:results_weak_drag_variations}

To further assess the impact of different atmospheric components, we apply simple modifications to the weak-drag model (i.e. the ``nominal'' model in Figure \ref{fig:ccf_panels1}) to see how the CCF maps of the planet change as a result of these modifications. 

In the second row of Figure \ref{fig:ccf_panels1}, we set the iron abundances to $\log(\text{VMR}) = -4.3$ throughout the entire atmosphere. This scenario mimics an atmosphere where iron is uniformly distributed (in terms of VMR) and where no condensation takes place on the nightside, regardless of the temperature. As demonstrated in Figure \ref{fig:ccf_panels1}, the CCF maps of the uniform-iron model are very similar to those of the nominal model. This implies that the signal of the full limb reveals virtually nothing about the exact distribution of iron across the atmosphere. That is, we cannot differentiate between a model with uniform abundances and a model with iron abundances that follow from chemical equilibrium. The CCF maps of the nominal model do not serve as a proof for iron condensation on the nightside of the planet.

In the third row of Figure \ref{fig:ccf_panels1}, we set all winds to zero, such that the only contribution to the Doppler shift comes from planetary rotation. The RV shifts of the individual sectors are constant during the transit (excluding ingress and egress), but the heights of the CCF peaks still change over time. This proves that the varying signal strength is \emph{not} an effect of atmospheric dynamics. As a result of the constant RV shifts in every sector, the CCF map of the full limb clearly features two modes. Both modes compete for the global maximum, which initially lies at positive RV. Halfway through the transit, however, it jumps to negative RV, owing to the fact that the signal strengths of the sectors change over time.

\begin{figure*}
\centering

\makebox[\textwidth][c]{\hspace{-10pt} \includegraphics[width=0.88\textwidth]{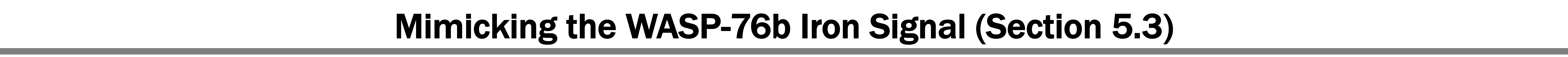}}

\vspace{-55pt}

\makebox[\textwidth][c]{ \vspace{-200pt} \includegraphics[width=1.2\textwidth]{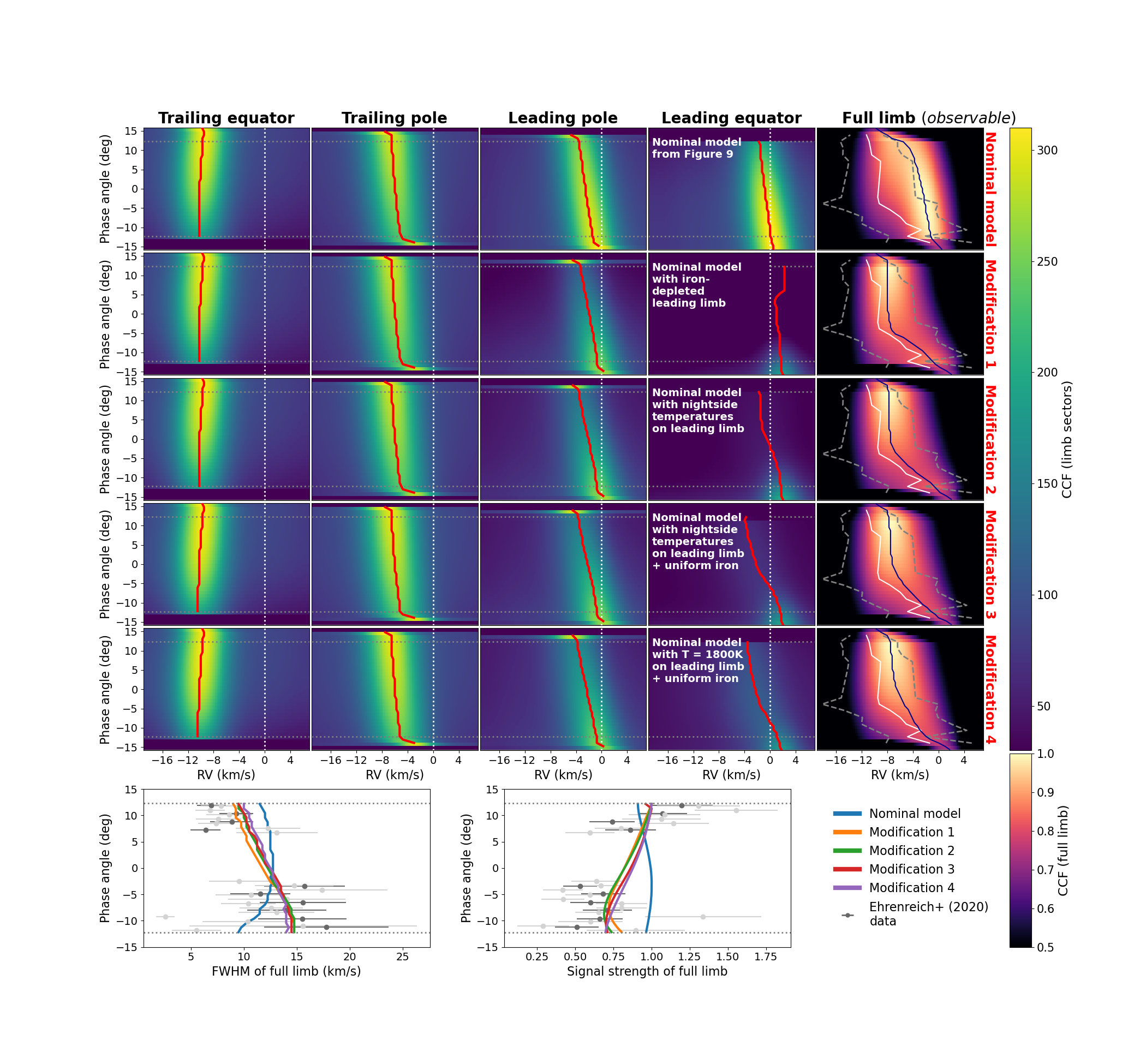}}

\vspace{-40pt}

\caption{Same as Figure \ref{fig:ccf_panels1}, but now the nominal model was changed in such a way that the CCF map of the full limb resembles the WASP-76b signal as observed by \citet{Ehrenreich2020}. In the panels in the right column, the white curve shows the peak position of the CCF that was obtained from VLT/ESPRESSO data, while the grey dashed lines indicate the corresponding FWHM. In the bottom panels, the FWHM and the signal strength of the models are plotted, along with data points from \citet{Ehrenreich2020}.}
\label{fig:ccf_panels2}
\end{figure*}

In the fourth row of Figure \ref{fig:ccf_panels1}, we set $T = 2300$ K in all cells and we use constant abundances throughout the entire atmosphere, again with $\log(\text{VMR}) = -4.3$ for iron. As in the third row, we only consider Doppler shifts due to rotation when simulating the transit spectra. The uniform-temperature model does not exhibit any longitudinal variations and so the CCF peak heights of the individual sectors remain constant during the transit. In conjunction with the other plots in Figure \ref{fig:ccf_panels1}, this result proves that the observed changes in the signal strengths of the sectors are due to the 3D temperature structure of the atmosphere. Because the signal strengths of the uniform-temperature model remain constant, the global maximum in the CCF map of the full limb does not make a jump around mid-transit. 


\subsection{Nightside Condensation of Iron in the Atmosphere of WASP-76b?} \label{ss:results_weak_drag_iron_removed}

The match between our nominal model from Figure \ref{fig:ccf_panels1} and the observations by \citet{Ehrenreich2020} and \citet{Kesseli2021} is not perfect yet. In the first half of the transit, the \emph{observed} iron signals (see Figure \ref{fig:ehrenreich_rv}) blueshift a lot faster than the signal that results from our weak-drag atmosphere. As suggested by \citet{Ehrenreich2020}, this discrepancy could be due to the fact that iron is in reality absent on the leading limb of WASP-76b. In what follows, our goal is (i) to check whether removing iron from the leading limb indeed results in a better match between the model and the data, and (ii) to explore whether the signal can be reproduced in any alternative way. To this end, we again apply a number of ad hoc modifications to the weak-drag model, which are summarised in Figure \ref{fig:modified_vmr}.

To test the hypothesis from \citet{Ehrenreich2020}, we take the nominal model from Figure \ref{fig:ccf_panels1} and we remove gaseous iron from all atmospheric cells with longitudes\footnote{An upper limit $\varphi = -58^\circ$ was found to give the best match to the data after some trial and error.} $-180^\circ < \varphi < -58^\circ$ and pressures $P < 0.01$ bar. The resulting CCF maps are shown in the second row of Figure \ref{fig:ccf_panels2} (modification 1). Modulo a small offset, the CCF map of the full limb provides a very good qualitative fit to the \emph{observed} iron signal. Both RV curves exhibit a steep slope in the first half of the observation, followed by a clear ``kink'' around mid-transit, after which the net RV shift remains virtually constant. 

The CCF maps of the individual limb sectors demonstrate how such an iron signal comes about. With respect to the nominal model, displayed in the top row, the maps of the trailing equator and the trailing pole are unaffected by the removal of iron. However, in the sectors on the leading limb, the iron signal fades away during the transit, as a larger part of the dayside rotates out of view. In fact, on the leading equator, the CCF peak has completely disappeared in the second half of the observation. As far as the CCF map of the full limb is concerned, this means that the trailing limb starts to dominate the average over time, such that the signal is pulled towards more negative RVs in the second half of the transit.

\begin{figure*}
\centering

\makebox[\textwidth][c]{\hspace{-10pt} \includegraphics[width=0.88\textwidth]{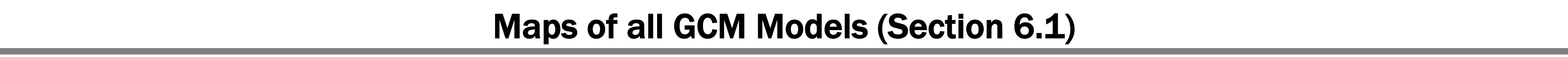}}

\vspace{-43pt}

\makebox[\textwidth][c]{ \vspace{-200pt} \includegraphics[width=1.2\textwidth]{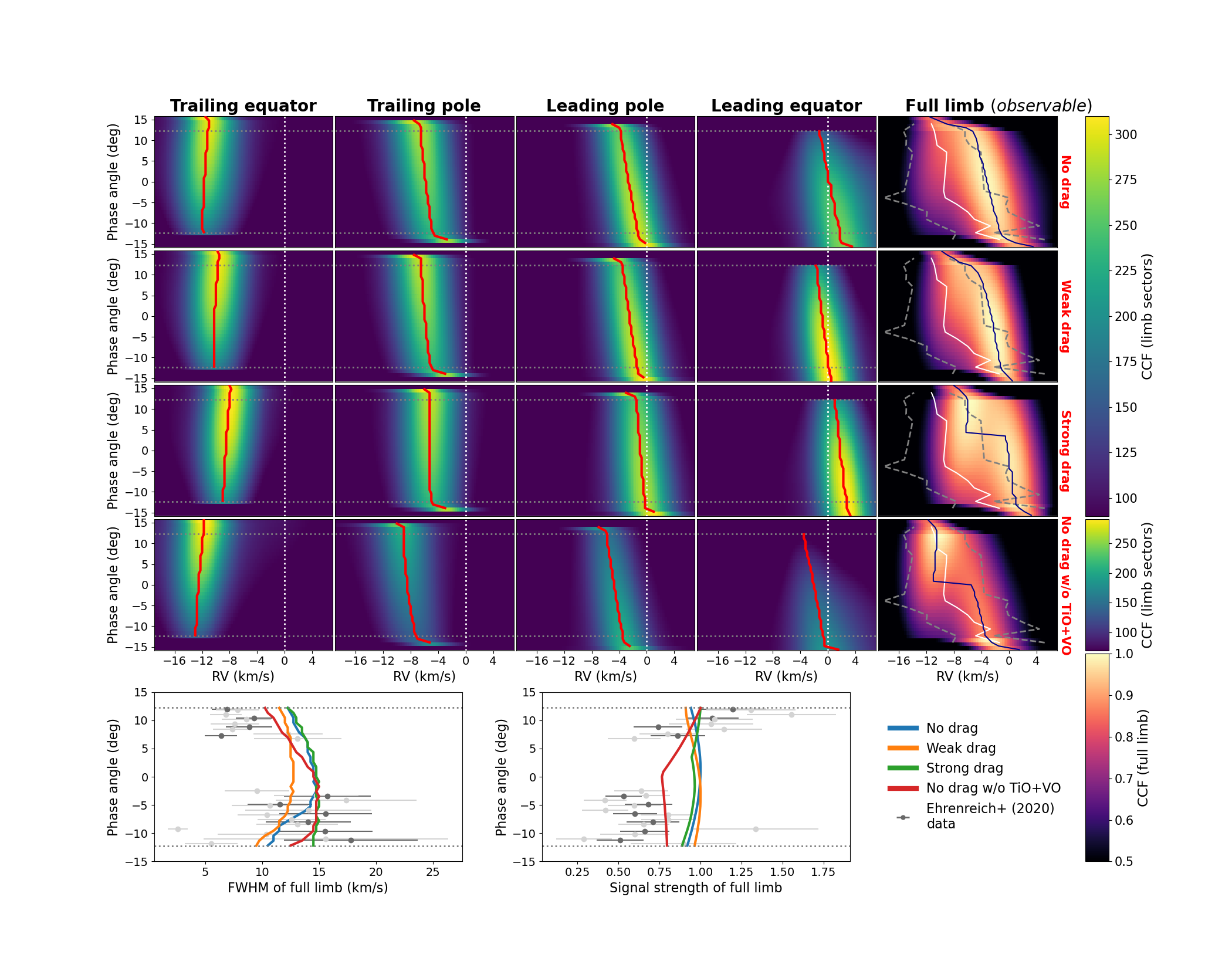}}

\vspace{-30pt}

\caption{Same as Figures \ref{fig:ccf_panels1} and \ref{fig:ccf_panels2}, but now for the four different GCM models. All maps account for Doppler shifts due to atmospheric dynamics and planetary rotation. Again, the iron signal observed by \citet{Ehrenreich2020} is plotted on top of the CCF maps of the full limb in the right column.}
\label{fig:ccf_panels3}
\end{figure*}

\begin{figure*}
\centering

\makebox[\textwidth][c]{\hspace{-10pt} \includegraphics[width=0.88\textwidth]{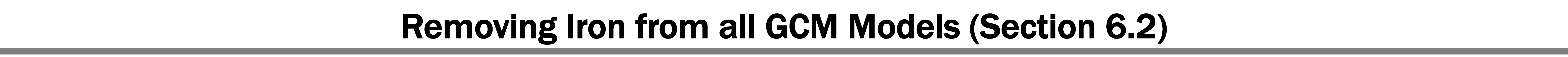}}

\vspace{-45pt}

\makebox[\textwidth][c]{ \vspace{-200pt} \includegraphics[width=1.2\textwidth]{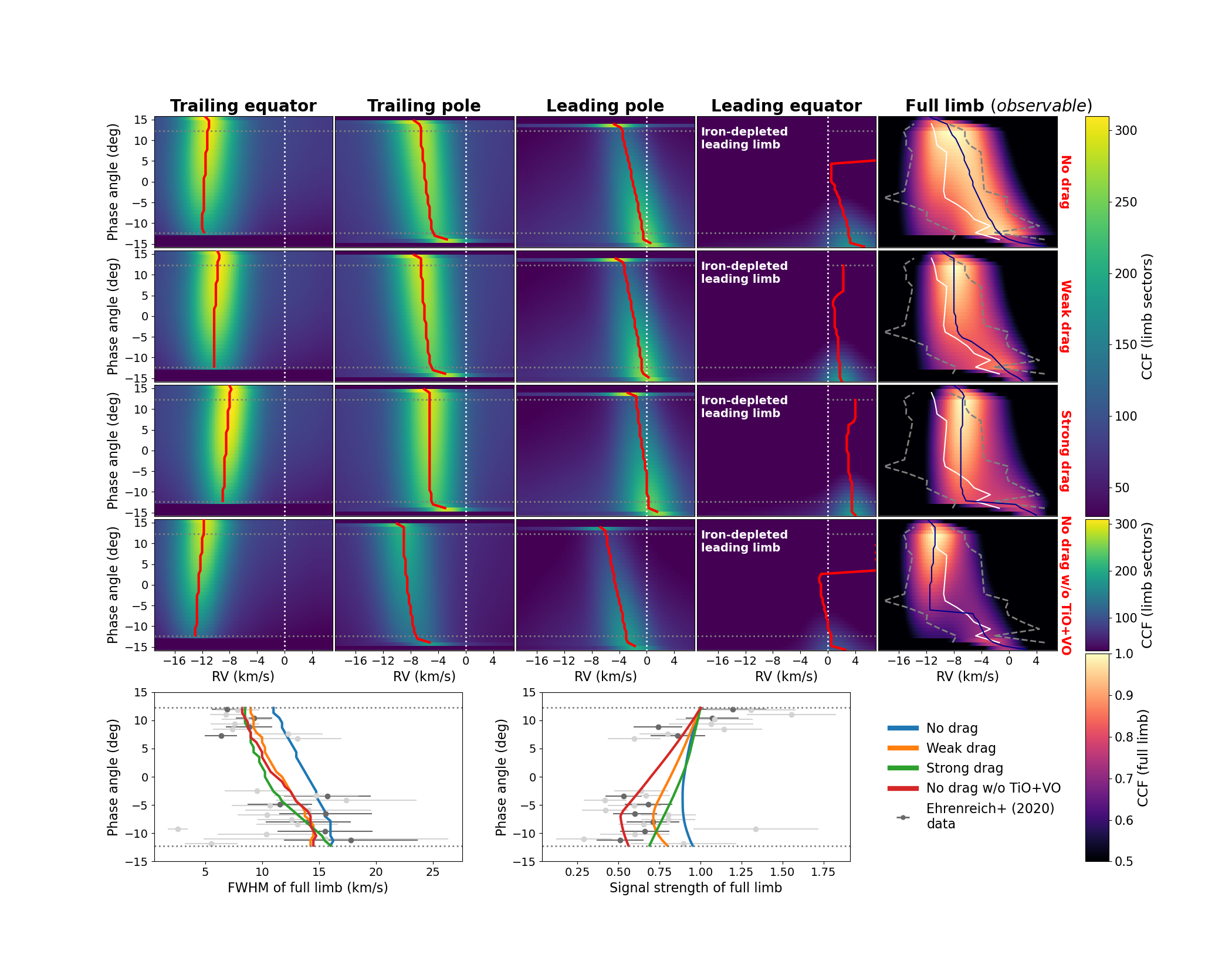}}

\vspace{-30pt}

\caption{Same as Figures \ref{fig:ccf_panels1}, \ref{fig:ccf_panels2} and \ref{fig:ccf_panels3}, but now for the four different GCM models with iron \emph{removed} from the leading limb. All maps account for Doppler shifts due to atmospheric dynamics and planetary rotation.}
\label{fig:ccf_panels4}
\end{figure*}

In the third row of Figure \ref{fig:ccf_panels2} (modification 2), we do not literally remove iron from the leading limb. Instead, we give all cells up to $-58^\circ$ longitude nightside temperatures, densities and abundances. As shown in Figure \ref{fig:modified_vmr}, this also results in a scale-height asymmetry between the trailing and the leading limb of the atmosphere. The motivation for this heuristic model are the results from the previous sections, which show that the temperature structure can impact the signal strength of individual limb sectors over time (see Figure \ref{fig:ccf_panels1}). This has a direct effect on the cross-correlation signal of the full limb. It turns out that the CCF maps resulting from nightside temperatures on the leading limb are nearly identical to those of the nominal model with iron removed.

In the fourth row of Figure \ref{fig:ccf_panels2} (modification 3), we take same model, but we additionally set the iron abundance to $\log(\text{VMR}) = -4.3$ across the \emph{entire} atmosphere (see the structure in Figure \ref{fig:modified_vmr}). In cells with $T < 1800$ K, we use the iron opacities at 1800 K, because iron would normally condense below this temperature. Remarkably, the CCF map of the full limb still provides an acceptable fit to the \citet{Ehrenreich2020} data, although the sharp kink around mid-transit has become less pronounced. This finding suggests that a substantial temperature difference between the planet's trailing and leading limb might be enough to reproduce the RV signal of \mbox{WASP-76b} -- regardless of whether iron condensation occurs. Of course, it is very likely that iron clouds form on the leading limb when when its temperature lies in the nightside regime. However, our simulations demonstrate that there could be (at least in theory) scenarios in which iron condensation is not strictly required to explain the planet's cross-correlation signal.

In the fifth row of Figure \ref{fig:ccf_panels2} (modification 4), we take our approach one step further. Instead of setting the temperatures on the leading limb equal to those on the nightside, we assume a constant temperature of 1800 K, such that gaseous iron could in reality persist in this region. As shown in the right panel in Figure \ref{fig:modified_vmr}, this also means that the leading limb is getting a larger scale height compared to the previous two models, such that its contribution to the total CCF will increase. As a result of these changes, the RV signal of the full limb loses its kink and the fit to the \citet{Ehrenreich2020} data becomes qualitatively worse. That said, the model still provides an insightful demonstration of the effect of temperature on the signal strength. Even though the trailing and leading equator have the same iron abundances, the signal of the latter remains much weaker because of its lower temperature. 

The bottom-left panel in Figure \ref{fig:ccf_panels2} shows the FWHM of the CCF maps of the full limb for the different modifications, as well as the nominal weak-drag model. The measurements from \citet{Ehrenreich2020} are plotted in the background. The data points come in two shades of grey, corresponding to different transit observations. Whereas the nominal model fails to match the data, the FWHM of the modified CCF maps seems to agree much better with the observations. In particular, the modified models correctly capture the \emph{decrease} in the FWHM over time, from $\sim$15 km/s at ingress to $\sim$10 km/s at egress. In terms of signal strength (bottom-right panel in Figure \ref{fig:ccf_panels2}), the modified models also outperform the nominal weak-drag model. Their signal strength after ingress is $\sim$75$\%$ of their signal strength before egress. This trend roughly matches the data, although the observations seem to suggest an even bigger change over time.


\section{CCF Maps of all GCM Atmospheres}
\label{s:results_all_gcm}

\subsection{Nominal Models}
\label{ss:results_all_gcm_nominal}

Figure \ref{fig:ccf_panels3} displays the CCF maps associated with all GCM models introduced in Section \ref{ss:methods_gcm}. The maps of the weak-drag model, which were discussed in Section \ref{ss:results_weak_drag_nominal}, can be found in the second row. Figure \ref{fig:ccf_panels3} illustrates that the drag timescale $\tau_{\text{drag}}$ of the atmosphere can have a large effect on cross-correlation signal of the full limb -- even though the changes to signals of the individual sectors may appear subtle.

The first row of Figure \ref{fig:ccf_panels3} shows the CCF maps of the GCM model without drag ($\tau_{\text{drag}} \rightarrow \infty$). When drag forces are absent, the atmosphere develops a super-rotating jet around the equator. The jet points towards the observer on the trailing limb and away from the observer on the leading limb of the planet. Compared to the weak-drag model, the RV signal of the trailing equator is further blueshifted by $\sim$1.5 km/s in the drag-free case. The RV signal of the leading equator has a steeper slope and is more redshifted. Also, its lower peak values are likely due to Doppler broadening, caused by a larger variance of the line-of-sight velocities within the sector (see Figure \ref{fig:gcm_terminator_vlos}). As a result of these changes, the RV signal of the full limb has a somewhat steeper slope compared to the weak-drag model. Nonetheless, it is questionable to what degree observations can really differentiate between both atmospheres -- that is, between a jet and a no-jet scenario. 

The third row of Figure \ref{fig:ccf_panels3} shows the CCF maps of the GCM model with strong drag ($\tau_{\text{drag}} = 10^4$ s). The shorter the drag timescale of the atmosphere, the further winds are slowed down. This effect is clearly visible on the trailing and leading equator, where the RV signals appear more redshifted compared to the weak-drag model. Reminiscent of the atmosphere in which the winds were set to zero (third row in Figure \ref{fig:ccf_panels1}), the CCF map of the full limb again features two modes. In the second half of the transit, the global maximum jumps from the marginally redshifted to the blueshifted mode. 

The fourth row of Figure \ref{fig:ccf_panels3} shows what happens to the CCF maps of the drag-free atmosphere when the opacities of TiO and VO are set to zero, both in the GCM and in \textsc{hires-mcrt}. As discussed in Section \ref{ss:methods_gcm}, this alters the radiation feedbacks in the GCM, and thereby the atmosphere's temperature structure and wind profile (see Figures \ref{fig:gcm_equator_temp_vmr}--\ref{fig:gcm_terminator_vlos}). The CCF maps illustrate that the day-to-night winds in the drag-free atmosphere become stronger upon the removal of TiO and VO, because the RV signals shift to the left in every sector. Also, it is striking to see that the signals of the polar sectors and the leading equator become a lot weaker compared to the other models. Consequently, the strongly blueshifted signal of the trailing equator dominates in the second half of the observation, such that the global maximum in the CCF map of the full limb jumps to $-12$ km/s around mid-transit. Considering all four GCM atmospheres, the model without TiO and VO provides the best fit to the iron signal of WASP-76b (ignoring the modified atmospheres presented in Section \ref{ss:results_weak_drag_iron_removed}). Although it does not manage to reproduce the kink in the RV signal, it does seem to give rise to the correct amount of blueshift. In the other GCM models, the blueshift is clearly too small, because the signal from the leading sectors does not fade quickly enough. Nonetheless, it remains an open question to what extent the absence of TiO and VO is a ``competitive'' explanation for the iron signal of WASP-76b. Recently, \citet{Edwards2020} claimed evidence for significant abundances of TiO and VO in the planet's atmosphere, based on low-resolution emission spectra obtained with Hubble/WFC3. On the other hand, transmission spectra taken with the same instrument did not show a clear sign of either species.  

\begin{figure*}
\centering

\vspace{-40pt}

\makebox[\textwidth][c]{ \vspace{-400pt} \includegraphics[width=1.2\textwidth]{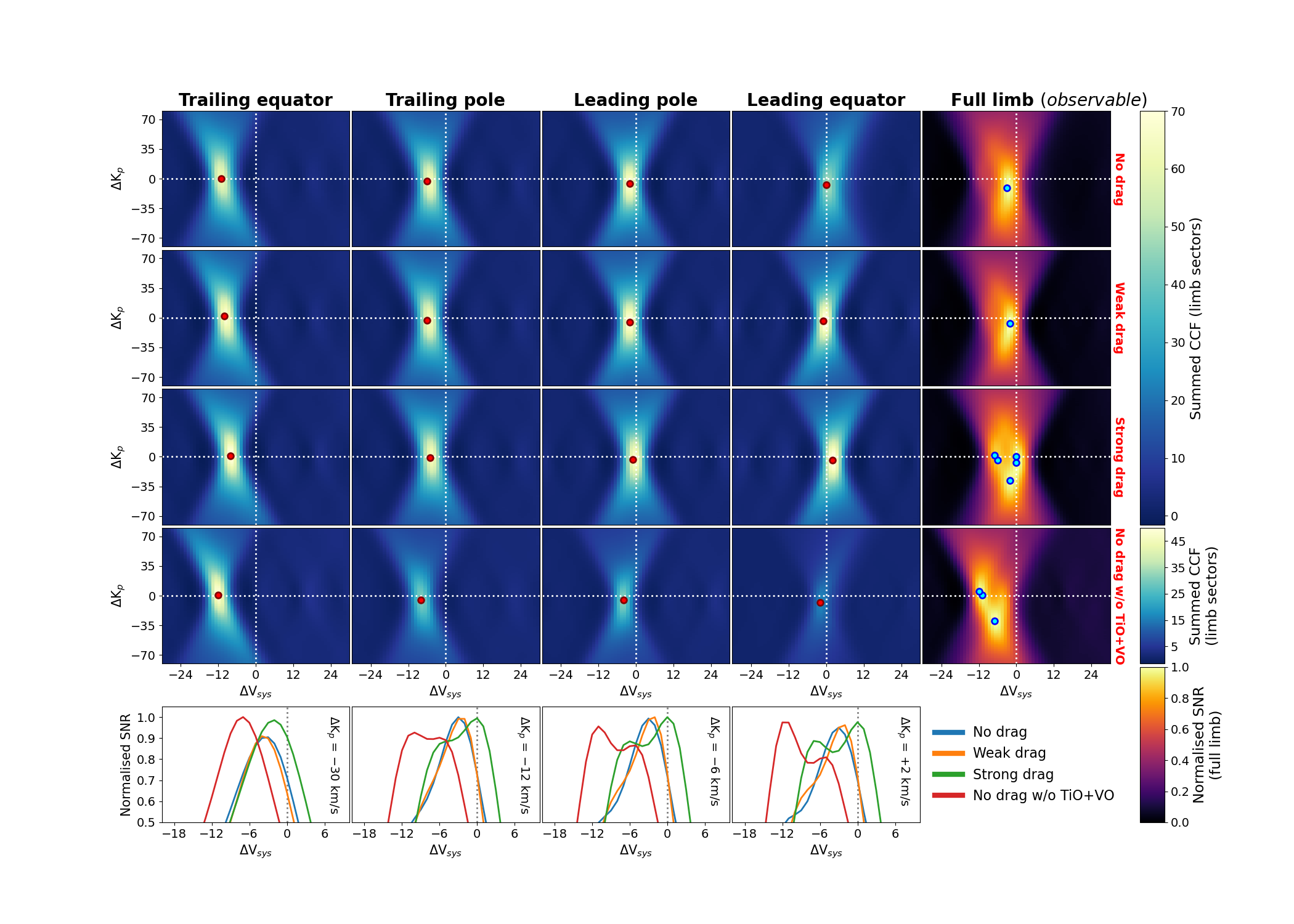}}

\vspace{-30pt}

\caption{\textbf{Top four rows:} $K_{\text{p}}$--$V_{\text{sys}}$ maps associated with the four GCM models from Figure \ref{fig:ccf_panels3}. In the panels pertaining to the limb sectors, we only summed (co-added) the CCF along the phase-axis, to provide a clear impression of a sector's contribution to the $K_{\text{p}}$--$V_{\text{sys}}$ map of the full limb. The SNR values of the full limb were normalised between 0 and 1 to allow for an easy comparison between the different models. The bullet markers in each of the panels indicate local maxima. \textbf{Bottom panels:} Cross-sections of the $K_{\text{p}}$--$V_{\text{sys}}$ maps of the full limb at four different values for $\Delta K_{\text{p}}$.}
\label{fig:kpvsys_normal}
\end{figure*}

With regards to Figure \ref{fig:ccf_panels3} as a whole, it is important to note that the RV signals of the individual limb sectors show a number of general trends across the four atmospheric models -- even though the CCF maps of the full limb may exhibit big differences. Firstly, all signals can be described by a straight line with a particular slope and offset. Also, the slopes have the same sign in each of the atmospheric scenarios. Furthermore, the signal strength of the trailing sectors always increases during the transit, while the strength of the leading sectors decreases over time. The rate of change is biggest in the equatorial regions, where cells move into (and out of) view the quickest. In other words, there is a correlation between a sector's rotational velocity and the change observed in the local temperature structure.

In the light of these general trends, it may be possible to parametrise the sector signals, and to retrieve them from the CCF map of the \emph{full limb} using an inverse-modelling technique. Although the feasibility of this method needs to be investigated, it could offer a unique way to extract spatial information from a planet's CCF map. Of course, success will also depend on the quality of the data (compare both panels in Figure \ref{fig:ehrenreich_rv}), but with a new generation telescopes and high-resolution instruments coming up (e.g. VLT/CRIRES+ and E-ELT/HIRES), future observations have the power to reveal unprecedented amounts of information.

\subsection{Removing Iron from the Leading Limb}
\label{ss:results_all_gcm_iron_removed}

By analogy with the approach in Section \ref{ss:results_weak_drag_iron_removed}, we now remove iron from the leading limb of the GCM atmospheres from Figure \ref{fig:ccf_panels3} (again up to $\varphi = -58^\circ$) to investigate how the structure of their RV signal changes. The results are presented in Figure \ref{fig:ccf_panels4}, where the second row pertains to the weak-drag model discussed previously.

\begin{figure*}
\centering

\vspace{-40pt}

\makebox[\textwidth][c]{ \vspace{-200pt} \includegraphics[width=1.2\textwidth]{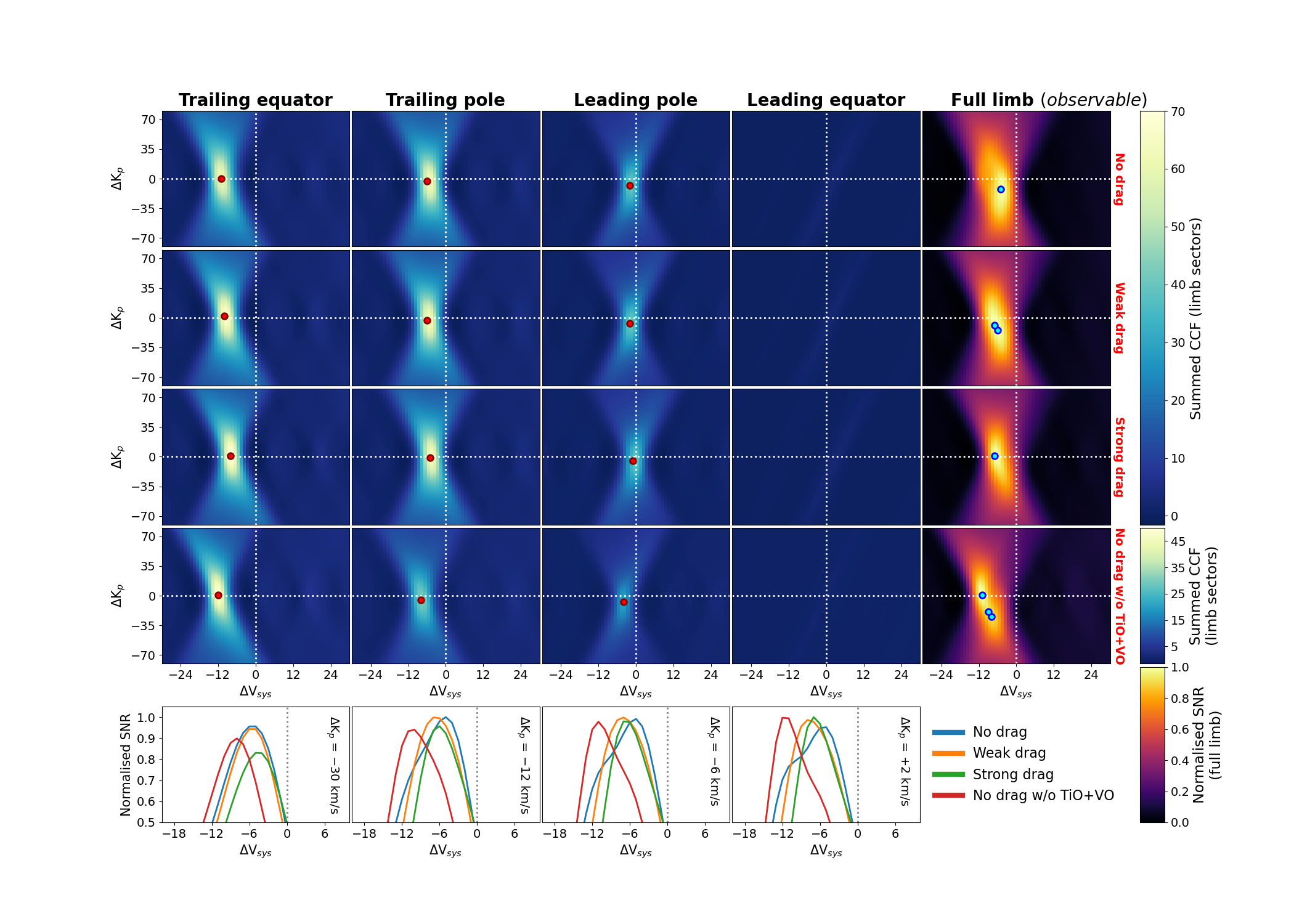}}

\vspace{-30pt}

\caption{Same as Figure \ref{fig:kpvsys_normal}, but now for the for the GCM models with iron removed from their leading limb. The corresponding CCF maps can be found in Figure \ref{fig:ccf_panels4}.}
\label{fig:kpvsys_modif}
\end{figure*}

The first row of Figure \ref{fig:ccf_panels4} shows the CCF maps of the drag-free model with iron removed from the leading limb. As far as the cross-correlations of the individual sectors are concerned, they appear very similar to those of the weak-drag model. However, the RV signals associated with the full limb show a clear discrepancy. Whereas the signal of the weak-drag model features a kink that is characteristic for the \citet{Ehrenreich2020} data, the RV slope of the drag-free model is more or less constant. A likely explanation for this is that the trailing equator of the drag-free model has a weaker signal compared to the weak-drag model, such that the impact of the trailing limb on the total average is smaller.

The third row of Figure \ref{fig:ccf_panels4} shows the CCF maps associated with the strong-drag atmosphere. In comparison to the weak-drag model, the CCF map of the full limb provides a worse qualitative fit to the observed WASP-76b iron signal. The RV signal acquires an \mbox{\texttt{L}-shape}, because the signals of the trailing sectors lie relatively close together. As a result, they start to dominate the full-limb average early on, especially while the signals from the leading sectors are fading away. 

The fourth row of Figure \ref{fig:ccf_panels4} shows the CCF maps of the drag-free model without TiO and VO opacities. The signal of the trailing limb is stronger than that of the other sectors and because the leading limb is depleted of iron, the RV signal of the full limb jumps to $-$12 km/s after a third of the transit. In terms of overall blueshift, the CCF map provides a good fit to the observed WASP-76b iron signal. However, in terms of the \emph{shape} of the RV signal, the fit is worse, because it features a jump rather than a kink.

\section{$K_{\text{p}}$--$V_{\text{sys}}$ Maps of all GCM Atmospheres}
\label{sect:kpvsys}

\subsection{Nominal Models}

Figure \ref{fig:kpvsys_normal} shows the $K_{\text{p}}$--$V_{\text{sys}}$ maps obtained for the four GCM models, whose CCF maps are displayed in Figure \ref{fig:ccf_panels3}. In this case, we only used the spectra from wavelength window 1 in Figure \ref{fig:cross_sections}, but we verified that the $K_{\text{p}}$--$V_{\text{sys}}$ maps associated with window 2 are similar. Because we performed all simulations in the planetary rest frame, one would naively expect the SNR to peak at $(\Delta V_{\text{sys}}, \Delta K_{\text{p}}) = (0,0)$ km/s. However, as the planet spectra are Doppler-shifted by winds and rotation, they may actually emulate the signal from a planet with a static atmosphere, but a nonzero system velocity and velocity semi-amplitude. As a consequence, the corresponding SNR maximum is displaced from $(0,0)$ km/s. In general, the value of $\Delta V_{\text{sys}}$ is set by the \emph{offset} of the RV signal in the cross-correlation map, while its \emph{slope} determines the value of $\Delta K_{\text{p}}$.

\begin{figure*}
\centering

\makebox[\textwidth][c]{ \hspace{-15pt} \includegraphics[width=1.2\textwidth]{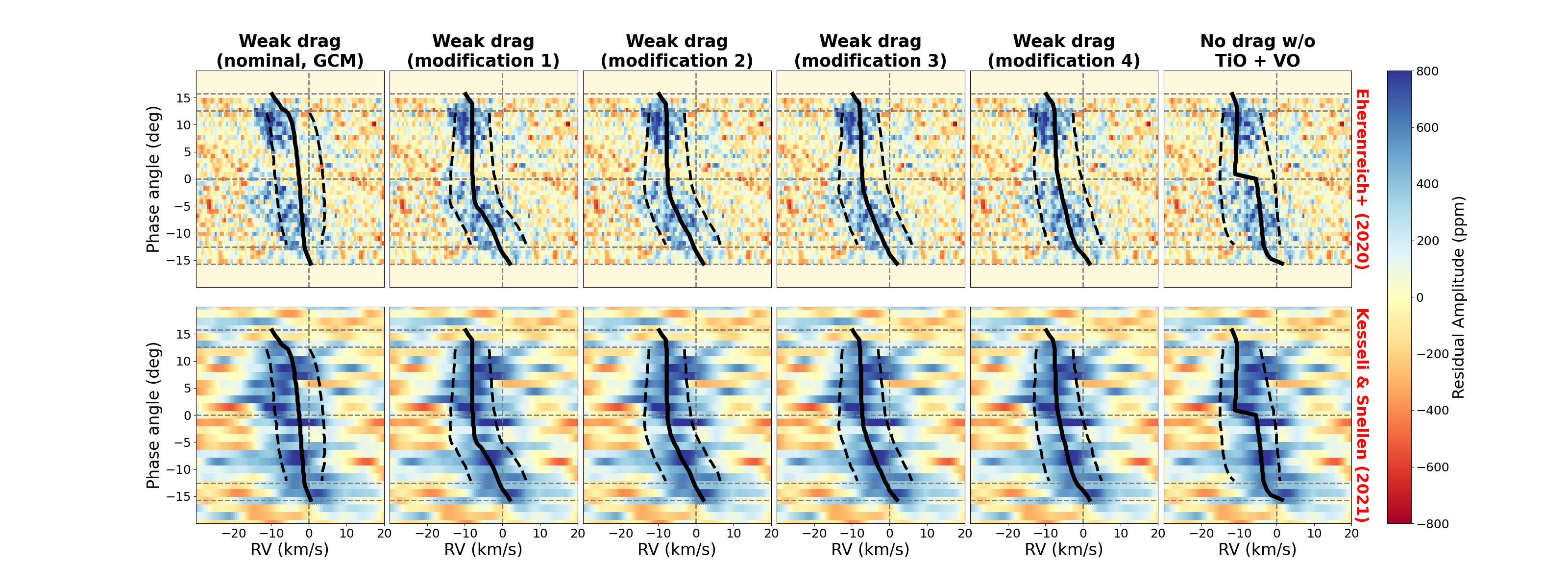}}

\vspace{-10pt}

\caption{The CCF maps from \citet{Ehrenreich2020} (top row) and \citet{Kesseli2021} (bottom row), with some of the models presented in this work plotted on top. The left column shows the nominal weak-drag model from the GCM (discussed in Section \ref{ss:results_weak_drag_nominal}), while the four central columns show the modified weak-drag models covered in Section \ref{ss:results_weak_drag_iron_removed}. As shown in Figure \ref{fig:modified_vmr}, Modification 1 has an iron-depleted leading limb, modification 2 has nightside temperatures on the leading limb, modification 3 has nightside temperatures on the leading limb and uniform iron abundances, while modification 4 has a leading limb of 1800 K and uniform iron abundances. The right column depicts the signal obtained from the drag-free atmosphere without TiO and VO opacities (see Section \ref{ss:results_all_gcm_nominal}). This is the GCM model that comes closest to the data without making any changes to the atmospheric structure.}
\label{fig:final_conclusion_figure}
\end{figure*}

The first four columns of Figure \ref{fig:kpvsys_normal} show the $K_{\text{p}}$--$V_{\text{sys}}$ maps associated with the individual limb sectors. The SNR was computed from Equation \ref{eq:def_kpvsys}, with $a = 1$ and $b = 0$, to allow for a comparison between the contributions from each of the sectors. The panels in the right column display the $K_{\text{p}}$--$V_{\text{sys}}$ maps of the full limb, with $a$ and $b$ chosen such that the SNR ranges from zero to unity in each map. Figure \ref{fig:kpvsys_normal} illustrates that all limb sectors have a single SNR maximum, regardless of the atmospheric scenario. This is because one can fit a straight line with a well-defined slope and offset to the RV signals of the limb sectors in Figure \ref{fig:ccf_panels3}. In accordance with the CCF maps, the SNR maxima of the trailing equator are most blueshifted (most negative $\Delta V_{\text{sys}}$), while the SNR maxima of the leading sectors are situated near $\Delta V_{\text{sys}} = 0$ km/s. Because the RV signals of the limb sectors only shift a couple of km/s during the transit, all SNR maxima lie close to $\Delta K_{\text{p}} = 0$ km/s. 

The $K_{\text{p}}$--$V_{\text{sys}}$ maps of the full limb are not necessarily single-peaked. Whereas the RV signals of the drag-free and the weak-drag model can be matched by a straight line (Figure \ref{fig:ccf_panels3}), the strong-drag model and the drag-free model without TiO and VO yield CCF maps with a more complicated structure. Their RV signals do not have a unique slope and offset, which leads to multiple maxima in \mbox{$K_{\text{p}}$--$V_{\text{sys}}$} space. Remarkably, this kind of behaviour was recently reported by \citet{Nugroho2020}, who found a ``double-peak'' feature in the Fe \textsc{i} signal of MASCARA-2b/KELT-20b. After a thorough investigation into its origin, they suggested it may be a product of atmospheric dynamics. Also, \citet{Nugroho2020} were able to generate a \mbox{$K_{\text{p}}$--$V_{\text{sys}}$} map with multiple peaks from a fabricated CCF map that contained two separate trails -- potentially analogous to separate contributions from a trailing and a leading limb.

Figure \ref{fig:kpvsys_normal} demonstrates that the locations of the SNR maxima of the full limb do not necessarily coincide with those of the individual limb sectors, as the sectors contribute only partially to the total signal. The map of the weak-drag model, for instance, looks diamond-shaped and features a local maximum at ($-2$, $-28$) km/s. The $K_{\text{p}}$--$V_{\text{sys}}$ map of the drag-free model without TiO and VO opacities acquires its highest value at ($-7$, $-30$) km/s. These results show how relatively small RV shifts in a CCF map can lead to $\Delta K_{\text{p}}$ offsets of tens of km/s. When the RV signal features a jump, the best-fitting straight line automatically gets a steeper slope, associated with a large $\Delta K_{\text{p}}$ value.

\subsection{Removing Iron from the Leading Limb}

Figure \ref{fig:kpvsys_modif} shows what happens to the $K_{\text{p}}$--$V_{\text{sys}}$ maps when iron is removed from the leading limb of the GCM atmospheres, just like in Sections \ref{ss:results_weak_drag_iron_removed} (modification 1) and \ref{ss:results_all_gcm_iron_removed}. Because the contributions from the leading pole and leading equator fade, the SNR landscape is dominated by the trailing sectors. For the drag-free and the weak-drag model, this means that the SNR maxima of the full limb show further blueshift, away from (0,0) km/s. Moreover, the signal of the strong-drag model loses its diamond shape, because the redshift contribution from the leading equator is now zero. In this respect, a diamond-shaped \mbox{$K_{\text{p}}$--$V_{\text{sys}}$} map could indicate a \emph{lack} of asymmetry between the trailing and the leading limb of a planet, although this needs to be verified more thoroughly. The maxima of the drag-free model without TiO and VO opacities apprear less affected by the removal of iron from the atmosphere. This is because even without iron depletion, the signal from the trailing equator is already a lot stronger than that of the leading equator. 

\section{Summary \& Conclusion}
\label{sec:conclusion}

Interpreting spectra that originate from inherently 3D exoplanet atmospheres is a challenge. This is especially true for ultra-hot Jupiters, which are characterised by an extreme day-night temperature contrast that drives a complex atmospheric circulation pattern. Because transmission spectroscopy probes the terminator of the atmosphere, it is sensitive to the regions where the largest thermal and chemical gradients occur. Since the transit of an ultra-hot Jupiter typically subtends a large angle ($\sim$30$^\circ$ in the case of WASP-76b), the high-resolution spectrum of the planet can undergo significant changes with time (see Figure \ref{fig:wasp76b_spectra}), and this complicates the cross-correlation signal in the planetary rest frame. Many observational studies have reported Doppler shifts in the absorption signals of ultra-hot Jupiters, but a clear picture of the dependencies is still lacking. This is why we require 3D frameworks to draw concrete links between (i) an atmosphere's structure, composition and dynamics, and (ii) the observed spectra.

In this work, we used the SPARC/MITgcm to simulate different atmospheric scenarios of the ultra-hot Jupiter WASP-76b (see Section \ref{ss:methods_gcm}). We considered models with three different drag timescales. Also, we investigated the effect of setting TiO and VO opacities to zero in the drag-free model. We then employed a Monte-Carlo radiative transfer code, \textsc{hires-mcrt}, to compute high-resolution transmission spectra with iron (\mbox{Fe \textsc{i}}) lines at 37 orbital phases during the transit (see Section \ref{ss:methods_mcrt}). \textsc{hires-mcrt} randomly draws transit chords along the line of sight, for which the optical depth is computed in a 3D spherical geometry. The transmission spectrum is then found by averaging over a sufficiently large sample of photon packets (see Equation \ref{eq:def_transit_area}). 

The (time-dependent) signal from the full limb of the planet can be hard to interpret due to the interplay between temperature, chemistry, dynamics and rotation. Hence, we decomposed the atmospheric annulus into four limb sectors (see Figure \ref{fig:limb_sectors_schematic}) and analysed the cross-correlation signals from each of the sectors separately. In the end, the cross-correlation map of the full limb is equal to the average of the maps of the individual sectors (see Equation \ref{eq:ccf_is_weighted_average}).

By applying simple modifications to the weak-drag model (see Figure \ref{fig:ccf_panels1}), we were able to identify the impact of the iron distribution and the temperature structure on the cross-correlation signal. We showed that the introduction of uniform iron abundances hardly alters the final observable. On the other hand, the vast temperature difference between the dayside and nightside causes the signal strengths of the sectors to change over time, and this affects the structure of the cross-correlation map. In addition, we investigated the nature of the WASP-76b iron signal observed by \citet{Ehrenreich2020} and \citet{Kesseli2021}. We demonstrated that the signal can be qualitatively reproduced by (i) removing gaseous iron from the leading limb of the weak-drag model and not changing the temperature structure, or (ii) applying a large temperature difference between the trailing and the leading limb of the planet. In case (ii), one might argue that a model without iron condensation provides a sufficiently good match to the data (see Figures \ref{fig:modified_vmr} and \ref{fig:ccf_panels2}), although iron condensation becomes inevitable when the temperature (and thereby the scale height) on the leading limb is further reduced. That said, the kink in the signal is most pronounced when iron is completely gone. Since our work was mainly exploratory, follow-up studies will have to shed light on the physical mechanisms that can give rise to (apparently large) asymmetries between the trailing and the leading limb of ultra-hot Jupiters. 

In Section \ref{ss:results_all_gcm_nominal}, we computed the cross-correlation maps of all four GCM atmospheres (see Figures \ref{fig:ccf_panels3} and \ref{fig:ccf_panels4}), and we showed that the drag timescale can have a substantial impact on the structure of the cross-correlation map. This is because the planet's wind profile governs the slopes and offsets of the RV signals originating from the limb sectors. We also showed that the cross-correlation map of the drag-free atmosphere changes drastically upon the removal of TiO and VO opacities, mainly because the new temperature structure affects the relative strength of the signals associated with individual sectors. Finally, the  $K_{\text{p}}$--$V_{\text{sys}}$ maps of the GCM atmospheres were presented in Section \ref{sect:kpvsys}, and we showed that atmospheric dynamics can give rise to multiple SNR maxima that are displaced from (0,0) km/s in the planetary rest frame.

Our most important conclusions are summarised below: 

\begin{enumerate}
   \setlength\itemsep{1em}
   \item[$\bullet$] The cross-correlation map of an ultra-hot Jupiter primarily encodes information about its temperature structure, dynamics and rotation. Based on the cross-correlation map, it is not possible to differentiate between (i) an atmosphere with uniform iron abundances and (ii) an atmosphere with equilibrium chemistry, where iron condenses at low (nightside) temperatures (see Figures \ref{fig:ccf_panels1} and \ref{fig:ccf_panels2}). Of course, this holds as long as the abundances on the trailing and the leading limb are high enough to be detected.
   
   \item[$\bullet$] The temperature contrast between the dayside and the nightside causes the signal strengths associated with limb sectors to change during the transit (see Figure \ref{fig:ccf_panels1}). The hotter a region probed by the observation, the stronger the iron signal becomes (see Figure \ref{fig:trailing_limb_spec}). Changes are biggest around the equator, where atmospheric regions move into (or out of) view the quickest. Due to these variations, the RV signal of an ultra-hot Jupiter may feature jumps or kinks (see e.g., Figures \ref{fig:ccf_panels3} and \ref{fig:ccf_panels4}). 
   
   \item[$\bullet$] Atmospheric drag can have a considerable impact on the cross-correlation signal of ultra-hot Jupiters, because it slows down winds and suppresses jet formation. In the regime of strong drag \mbox{($\tau_{\text{drag}} \sim 10^4$ s)}, the cross-correlation signal features two modes (see Figure \ref{fig:ccf_panels3}) that cause the global maximum to jump from 0 to $-6$ km/s after mid-transit.  
   
  \item[$\bullet$] Removing TiO and VO opacities from the drag-free model drastically alters the temperature structure of the atmosphere. The hotspot shift on the dayside extends to higher altitudes, such that the temperature (and scale-height) asymmetry between the trailing and leading limb is larger than in the other models (see Figure \ref{fig:gcm_equator_temp_vmr}). This causes the signal of the hotter, blueshifted trailing equator to be much stronger than the signals of the other limb sectors. 
  
  \item[$\bullet$] The iron signal of WASP-76b reported by \citet{Ehrenreich2020} and \citet{Kesseli2021} can be reproduced by removing iron from the leading limb of the weak-drag model, up to a longitude of $\phi \approx -58^\circ$. Additionally, the signal can be reasonably well matched by models with uniform iron abundances and a large temperature difference between the trailing and the leading limb (see Figures \ref{fig:modified_vmr} and \ref{fig:ccf_panels2}). In these scenarios, iron condensation (although likely to happen) is not strictly required.
  
  \item[$\bullet$] When iron is removed from the leading limb, the weak-drag model \mbox{($\tau_{\text{drag}} \sim 10^5$ s)} provides a better fit to the WASP-76b observations than the other GCM atmospheres (see Figure \ref{fig:ccf_panels4}). In fact, it is the only model that can reproduce the kink in the RV signal.
  
  \item[$\bullet$] Atmospheric dynamics and rotation can cause the peak in the $K_{\text{p}}$--$V_{\text{sys}}$ map to be substantially displaced from $(0,0)$ km/s in the planetary rest frame, especially along the $K_{\text{p}}$ axis (see Figures \ref{fig:kpvsys_normal} and \ref{fig:kpvsys_modif}). Additionally, there can be multiple (local) SNR maxima when the RV signal of the planet deviates too much from a single straight line. The $K_{\text{p}}$--$V_{\text{sys}}$ map of the strong-drag model, for instance, exhibits a diamond shape.
  
\end{enumerate}

\section*{Acknowledgements}

We are grateful to David Ehrenreich and Aurora Kesseli for providing the data of their CCF maps. We also thank Stevanus Nugroho and Neale Gibson for enlightening discussions, and Raymond Pierrehumbert for sharing computing resources. JPW sincerely acknowledges support from the Wolfson Harrison UK Research Council Physics Scholarship and the Science and Technology Facilities Council (STFC). Furthermore, JPW is much indebted to the staff of Wolfson College, Oxford, for their great efforts to keep services running during the COVID-19 pandemic. Finally, we thank the anonymous referee for thoughtful comments that helped improve the the quality of this manuscript.

\section*{Data Availability}

The data and models underlying this article will be shared on reasonable request to the corresponding author.



\bibliographystyle{mnras}
\bibliography{manuscript} 




\appendix

\section{Benchmarking hires-mcrt against Chimera}
\label{ap:B}

\begin{figure*}
\centering

\includegraphics[width=0.7\textwidth]{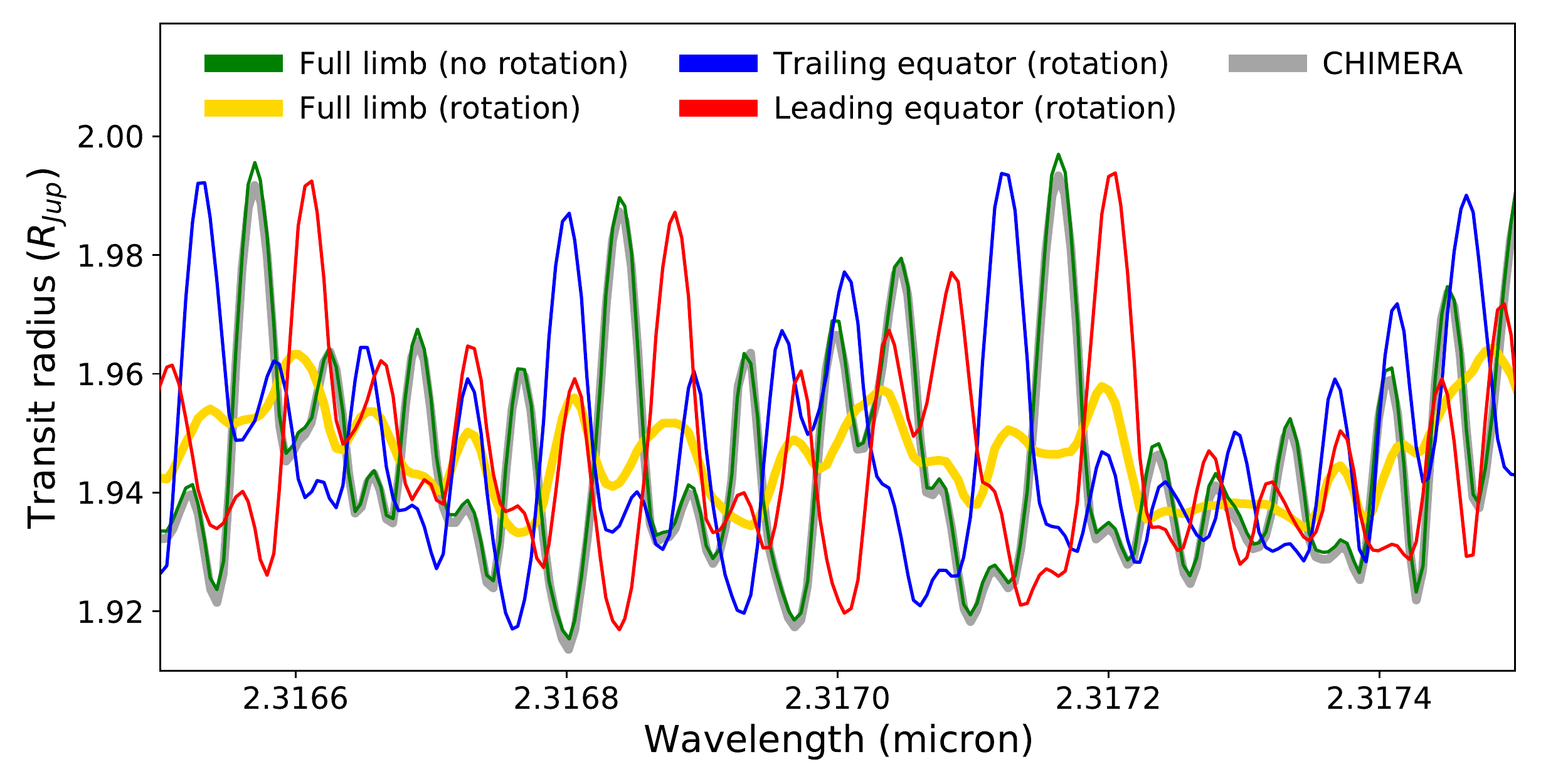}

\caption{Transmission spectra of the uniform atmosphere described in Appendix \ref{ap:B}. The green curve shows the spectrum of the static atmosphere \mbox{($v_{\textsc{los}} = 0$ km/s)}, while the yellow curve denotes the spectrum obtained when Doppler shifts due to rotation are taken into account. In the latter case, the red and blue curves represent the contributions from the trailing and leading equator (see Figure \ref{fig:limb_sectors_schematic}), respectively. The spectrum computed with the 1D radiative transfer code CHIMERA is plotted in grey.}
\label{fig:benchmark}
\end{figure*}

In this appendix, we aim to demonstrate that \textsc{hires-mcrt} works correctly in the limit of a uniform, 1D atmosphere. Since \citet{Lee2019} already subjected the Monte-Carlo code to extensive benchmark tests in low resolution, our only task here is to show that the high-resolution opacity tables are properly evaluated and that the Doppler shifts resulting from Equation \ref{eq:vlos} are correctly implemented. To this end, we generate a simple model atmosphere with 16 latitudes, 32 longitudes and 100 vertical levels, containing H$_2$, He, H$_\text{2}$O and CO. We adopt uniform number fractions $n_{\text{H2}} = 0.86$, $n_{\text{He}} = 0.14$, $n_{\text{H2O}} = 0.001$ and $n_{\text{CO}} = 0.01$, and we use a mean-molecular weight of 2.3 $m_{\text{H}}$. The atmosphere extends from $P$ = 200 bar (bottom) to \mbox{$P$ = $10^{-8}$} bar (top) and has a uniform temperature of 2000 K. Furthermore, $R_{\text{0}} = 1.76 R_{\text{Jup}}$ and $R_{\text{0}} + z_{\text{max}} = 2.05 R_{\text{Jup}}$. Also, we assume that the planet has the same orbital period as WASP-76b (1.81 days), such that the rotational velocity at the equator is $\pm 5.3$ km/s.


As shown in Figure \ref{fig:benchmark}, we compute a number of spectra with \textsc{hires-mcrt} in a very narrow region near 2.3 micron. Additionally, we employ the 1D, plane-parallel radiative transfer code CHIMERA (\citealt{Line2013, Line2014}) to compute the spectrum of the associated 1D atmosphere. In the radiative transfer, we account for the opacities of H$_2$O (\citealt{Polyansky2018}) and CO (\citealt{Li2015}), taken from the \texttt{ExoMol} database (\citealt{Tennyson2020}), as well the continuum due H$_2$-H$_2$ and H$_2$-He collision-induced absorption (\citealt{Borysow2001, Borysow2002, Gordon2017}). Figure \ref{fig:benchmark} demonstrates that there is a good agreement between the static Monte-Carlo spectrum, with $v_{\textsc{los}} = 0$ km/s in every cell, and the spectrum obtained from CHIMERA. Small differences are likely due to the way in which (the altitudes of) atmospheric levels are defined in the respective codes.

The transmission spectrum dramatically changes once \textsc{hires-mcrt} accounts for the Doppler shifts due to planetary rotation. In this case, the spectrum can be interpreted as an average over the four limb sectors (Figure \ref{fig:limb_sectors_schematic}), each with their own effective velocity shift and line widths. The contributions from the trailing and leading equator are shown in Figure \ref{fig:benchmark} as well. The trailing equator is blueshifted, because it rotates towards the observer. On the other hand, the leading equator is redshifted because it rotates away. The lines in both spectra are slightly less sharp compared to those of the static atmosphere, for there exists a slight dispersion in the line-of-sight velocities across the sectors.

Cross-correlating the equatorial spectra with the spectrum of the static atmosphere produces velocity shifts of $\pm 5.1$ km/s, which is in agreement with the orbital period of the planet. Note that one would expect these values to be slightly smaller than $\pm 5.3$ km/s, because the equatorial sectors defined in this work encompass all latitudes between $-45$ and $+45$ degrees.

\section{The line-of-sight velocity}
\label{ap:A}

\begin{figure}
\centering
\includegraphics[width=0.50\textwidth]{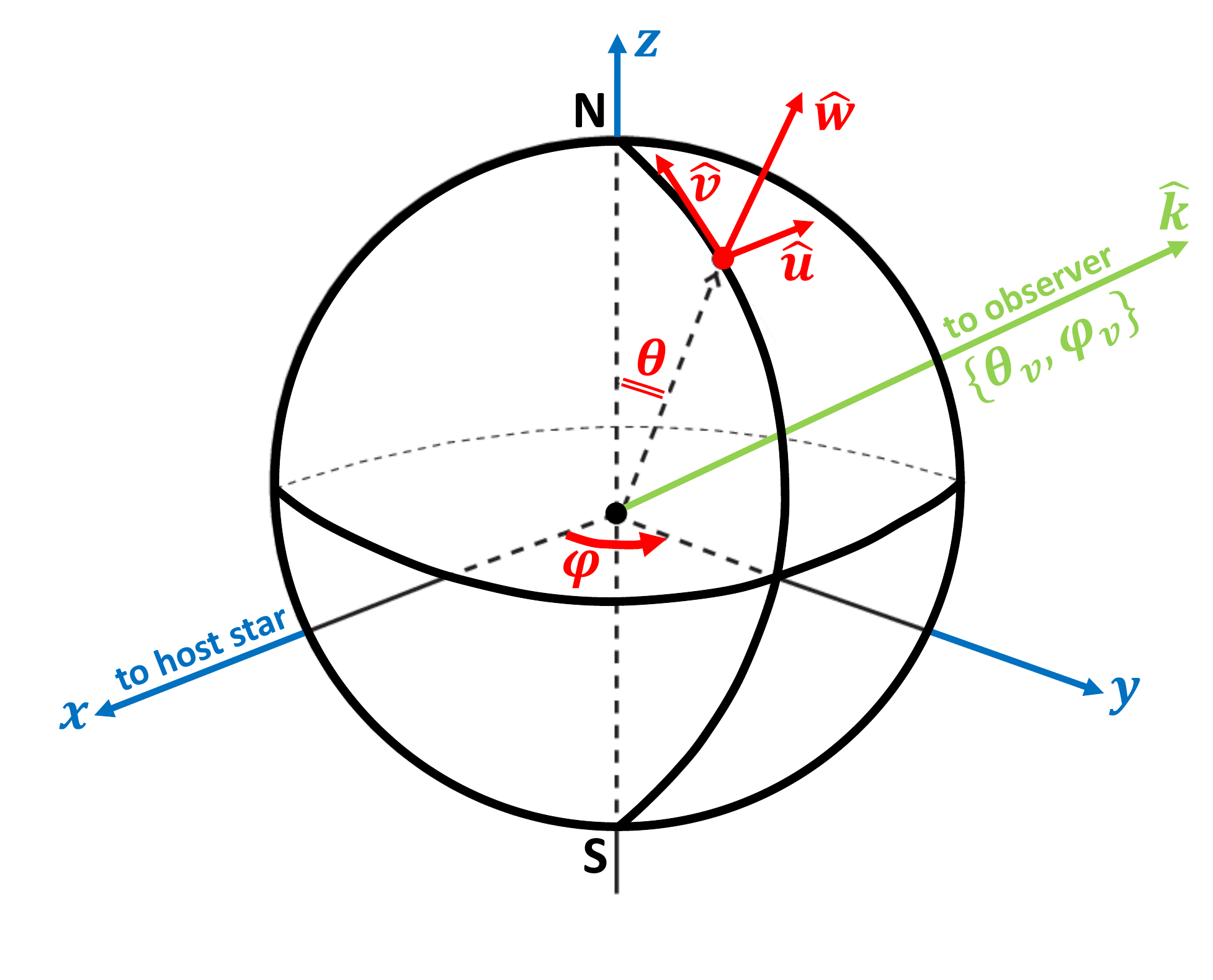}
\caption{A graphic depiction of a local coordinate system $\{\hat{u}, \hat{v}, \hat{w} \}$ (in red) and the frame that co-rotates with the planet (in blue). The angle $\varphi$ is measured counter-clockwise from the positive $x$ axis, while  $\theta$ is the angle between $\hat{w}$ and the $z$ axis. The viewing vector $\hat{k}$ is a unit vector which points towards the observer and has associated angles $\theta_v$ and $\varphi_v$. \texttt{N} and \texttt{S} denote the planet's north and south pole, respectively. Image adapted from \citet{Constantin2017}.}
\label{fig:sphere_coords}
\end{figure}

In this appendix, we will provide a derivation of Equation \ref{eq:vlos}, which we use to calculate the line-of-sight velocity by which the opacities are shifted in the Monte-Carlo code. We start by defining a local Cartesian coordinate system $\{\hat{u}, \hat{v}, \hat{w} \}$ in each atmospheric cell. As shown in Figure \ref{fig:sphere_coords}, $\hat{u}$ points into the local west-to-east direction, $\hat{v}$ points into the local south-to-north direction and $\hat{w}$ points upwards, perpendicular to the local surface.

In the atmospheric structure read by the Monte-Carlo code, the wind vectors are defined in the local frames of the atmospheric cells. Hence, to obtain the effective line-of-sight velocities, we must project the wind vectors onto a viewing vector $\hat{k}$, a unit vector that points towards the observer. It is most convenient to perform this projection in a coordinate system $\{\hat{x}, \hat{y}, \hat{z} \}$ that co-rotates with the planet (see Figure \ref{fig:sphere_coords}). Here, $\hat{x}$ and $\hat{y}$ span the orbital plane, such $\hat{x}$ always points towards the host star. In this setup, observing the planet at different phase angles or orbital inclinations is simply a matter of changing the viewing vector $\hat{k}$.

Our first step is to express the unit vectors $\{\hat{u}, \hat{v}, \hat{w} \}$ in the $\{\hat{x}, \hat{y}, \hat{z} \}$ frame:

\begin{equation} \label{eq:uvw}
\hat{u} = \begin{pmatrix} -\sin(\varphi)  \\ \cos(\varphi) \\ 0  \end{pmatrix}  \ \ \   \hat{v} = \begin{pmatrix} -\cos(\theta)\cos(\varphi)  \\ -\cos(\theta)\sin(\varphi) \\ \sin(\theta)  \end{pmatrix} \ \ \ \hat{w} = \begin{pmatrix} \sin(\theta)\cos(\varphi)  \\ \sin(\theta)\sin(\varphi) \\ \cos(\theta)  \end{pmatrix},
\end{equation}

\noindent where $\theta$ and $\varphi$ are the spherical coordinates of the cell, as denoted in Figure \ref{fig:sphere_coords}. The expression for $\hat{w}$ is trivial, as it points into the radial direction. Additionally, $\hat{u}$ is always tangent to the equator, so it must be confined to the $x$-$y$ plane. The expression for $\hat{v}$ is less intuitive, but can be obtained from the cross product $\hat{w} \times \hat{u}$. Additionally, we can write the viewing vector $\hat{k}$ as

\begin{equation} \label{eq:k}
\hat{k} = \begin{pmatrix} \sin(\theta_v)\cos(\varphi_v)  \\ \sin(\theta_v)\sin(\varphi_v) \\ \cos(\theta_v)  \end{pmatrix},
\end{equation}

\noindent with $\theta_v$ and $\varphi_v$ the viewing angles of the observer that define the line of sight. For an edge-on orbit, $\theta_v = 90^\circ$ and the third component of $\hat{k}$ becomes zero.

\subsection*{Contribution from Winds}

Given the wind vector $(u, v, w)$ in a particular atmospheric cell, the effective wind speed $v_{\textsc{los}}$ along the line of sight is given by

\begin{equation} \label{eq:vlos_appendix}
v_{\textsc{los}} = - \hat{k} \cdot \big( u \hat{u} + v \hat{v} + w \hat{w} \big),
\end{equation}

\noindent where the minus sign accounts for the fact that velocities are negative when they point towards the observer. Plugging in Equations \ref{eq:uvw} and \ref{eq:k}, and working out the inner products yields:

\begin{align}
\begin{split} 
    - u\hat{k} \cdot \hat{u} ={} & u \sin(\theta_v) \sin(\varphi - \varphi_v) \\
    - v\hat{k} \cdot \hat{v} ={} & v\cos(\theta)\sin(\theta_v) \cos(\varphi-\varphi_v) - v\sin(\theta)\cos(\theta_v) \\
    - w\hat{k} \cdot \hat{w} ={} & -w\sin(\theta)\sin(\theta_v)\cos(\varphi-\varphi_v) - w\cos(\theta)\cos(\theta_v)
\end{split}
\end{align}

\noindent In its full glory, Equation \ref{eq:vlos_appendix} now reads:

\begin{align}
\begin{split} \label{eq:app_res1}
    v_{\textsc{los}} ={} &  u \sin(\theta_v) \sin(\varphi - \varphi_v) \\
         & + \big[v\cos(\theta) - w\sin(\theta) \big] \sin(\theta_v)\cos(\varphi-\varphi_v) \\
         & - \big[ v\sin(\theta) + w\cos(\theta) \big] \cos(\theta_v).
\end{split}
\end{align}

\noindent To obtain the first three terms of Equation \ref{eq:vlos}, we need to make a number of substitutions. Firstly, $\theta_v = 90^\circ$ under the assumption of an edge-on orbit, such that the final term in Equation \ref{eq:app_res1} drops out. Furthermore, $\varphi_v = (180^\circ - \phi)$ and $\theta = (90^\circ - \alpha)$ allow for the line-of-sight velocity to be expressed in terms of the orbital phase angle $\phi$ and the cell's latitude $\alpha$. Note that these relationships change when the orientation of the orbit is different.

\subsection*{Contribution from Planetary Rotation}

The rotation of the planet also gives rise to a line-of-sight velocity term. This is because the $\{\hat{x}, \hat{y}, \hat{z} \}$ frame is rotating with respect to the observer. For a planet that is tidally locked, the rotation period is equal to the orbital period, with an associated angular frequency $\Omega$ (with units rad/s).

For each atmospheric cell, the velocity vector $\vec{\nu}$ due to rotation only points into the local $\hat{u}$ direction, perpendicular to the planetary axis. Its magnitude depends on the latitude and altitude $z$ of the atmospheric cell:

\begin{equation}
\vec{\nu} = \Omega (R_{\text{0}}+z) \sin(\theta) \hat{u}.
\end{equation}

\noindent Here, $R_{\text{0}}$ is the planetary radius at the bottom of the atmosphere, such that $R_{\text{0}}+z$ is the distance from the cell to the centre of the planet. $\Omega$ is positive when the $\{\hat{x}, \hat{y}, \hat{z} \}$ frame rotates counter-clockwise as seen from the positive $z$ axis. The resulting line-of-sight velocity is

\begin{align}
\begin{split} \label{eq:app_res2}
    v_{\textsc{los}} ={} &  - \hat{k} \cdot \vec{\nu} \\
            ={}& \Omega (R_0+z) \sin(\theta) \sin(\theta_v) \big[ \sin(\varphi)\cos(\varphi_v)- \cos(\varphi)\sin(\varphi_v) \big] \\
     ={} & \Omega (R_0+z) \sin(\theta) \sin(\theta_v) \sin(\varphi - \varphi_v)
\end{split}
\end{align}

\noindent By substituting $\theta_v = 90^\circ$, $\varphi_v = (180^\circ - \phi)$ and $\theta = (90^\circ - \alpha)$, we recover the final term in Equation \ref{eq:vlos}.

\section{Supplementary Figures}
\label{ap:C}

The supplementary figures in this appendix show the temperature, iron abundances and line-of-sight velocities in the limb plane of the four GCM models -- at ingress ($\phi=-14^\circ$), mid-transit ($\phi=0^\circ$) and egress ($\phi=+14^\circ$). The limb plane crosses the centre of the planet and is orthogonal to the line of sight. It coincides with the terminator plane at mid-transit.

\begin{figure*}
\centering

\vspace{-20pt}

\makebox[\textwidth][c]{\hspace{-30pt} \includegraphics[width=1.0\textwidth]{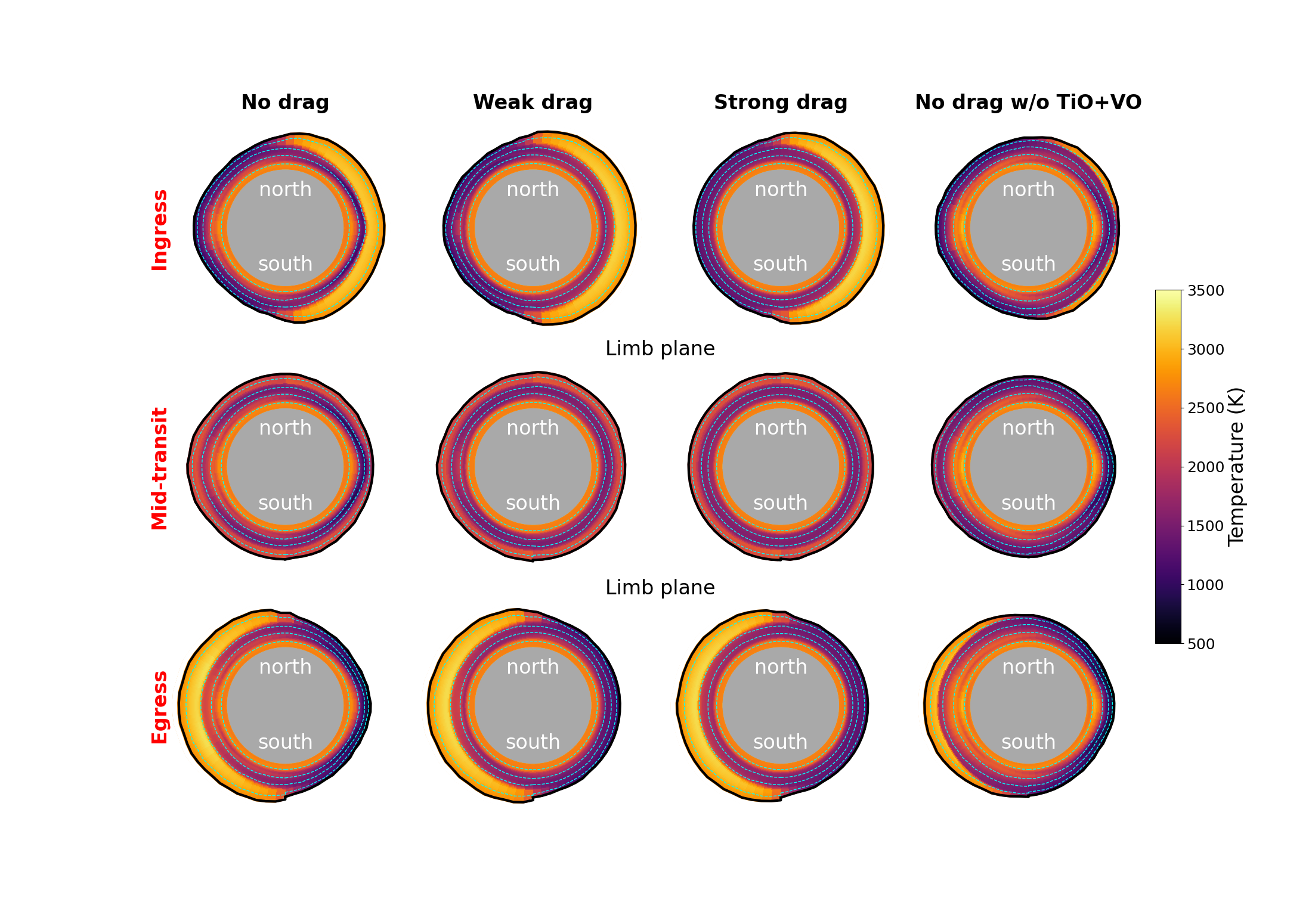}}

\vspace{-41pt}

\caption{Temperatures in the limb plane of the four GCM models (columns), at different phases during the transit (rows). The maps are plotted as a function of spatial coordinate, with the dashed lines denoting isobars of $10^1$, $10^{-1}$, $10^{-3}$ and $10^{-5}$ bar, respectively. The solid black contour marks the GCM boundary at 2 $\mu$bar.}
\label{fig:app_c1}
\end{figure*}

\begin{figure*}
\centering

\vspace{-20pt}

\makebox[\textwidth][c]{\hspace{-30pt} \includegraphics[width=1.0\textwidth]{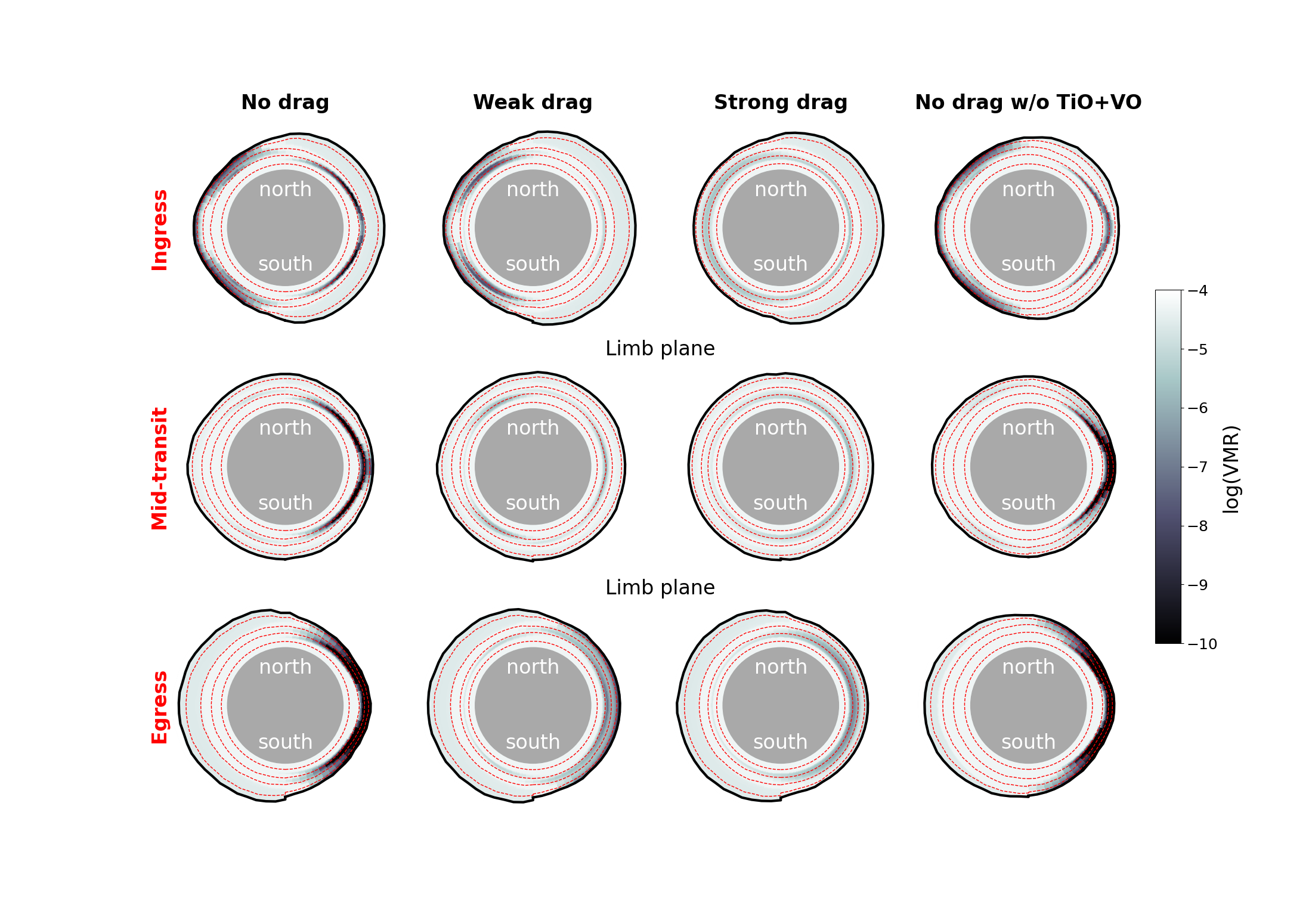}}

\vspace{-41pt}

\caption{Same as Figure \ref{fig:app_c1}, but now for the iron (Fe \textsc{i}) abundances.}
\label{fig:app_c2}
\end{figure*}

\begin{figure*}
\centering

\vspace{-20pt}

\makebox[\textwidth][c]{\hspace{-30pt} \includegraphics[width=1.0\textwidth]{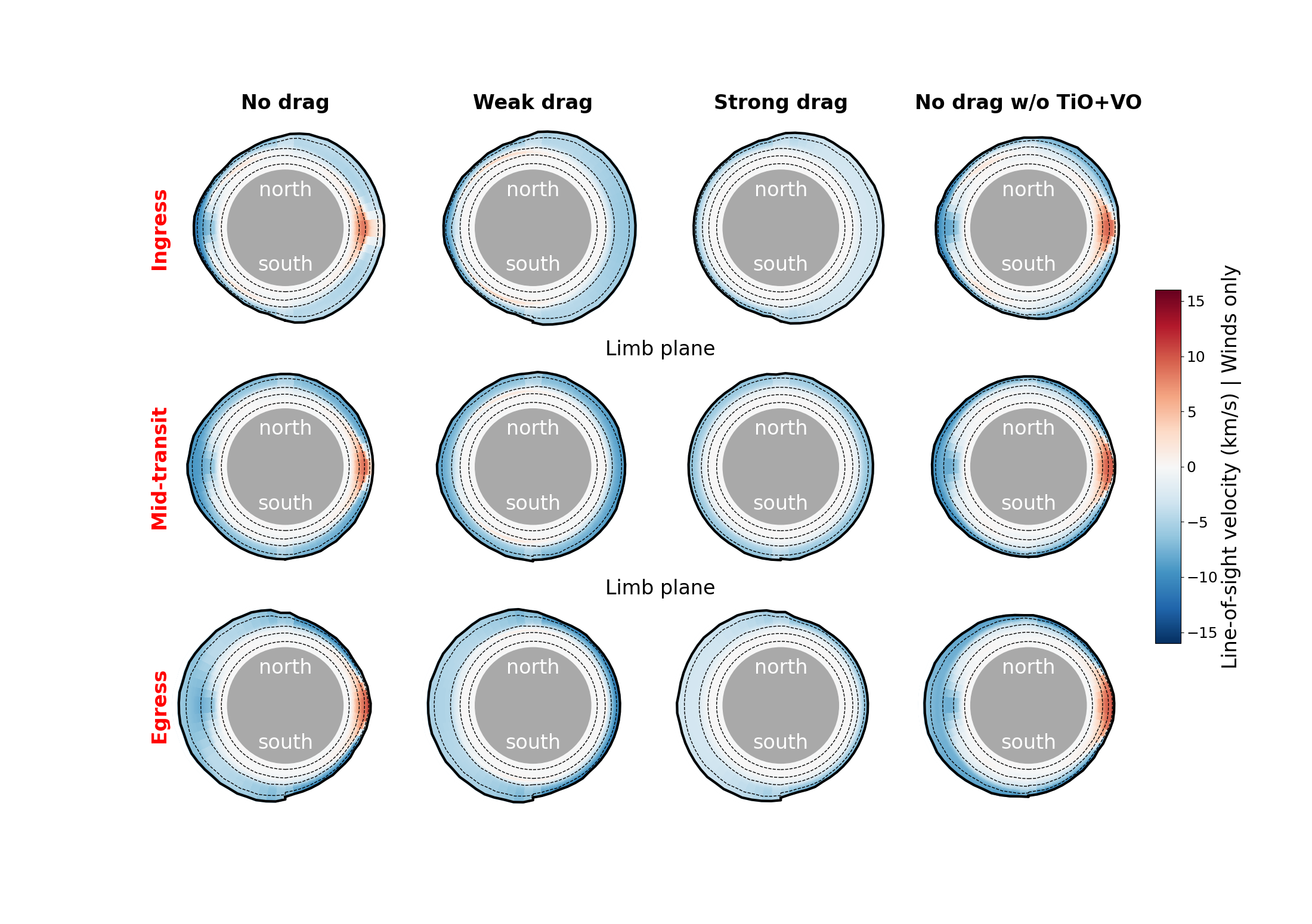}}

\vspace{-41pt}

\caption{Same as Figure \ref{fig:app_c1}, but now for the line-of-sight velocity due to winds only.}
\label{fig:app_c3}
\end{figure*}

\begin{figure*}
\centering

\vspace{-20pt}

\makebox[\textwidth][c]{\hspace{-30pt} \includegraphics[width=1.0\textwidth]{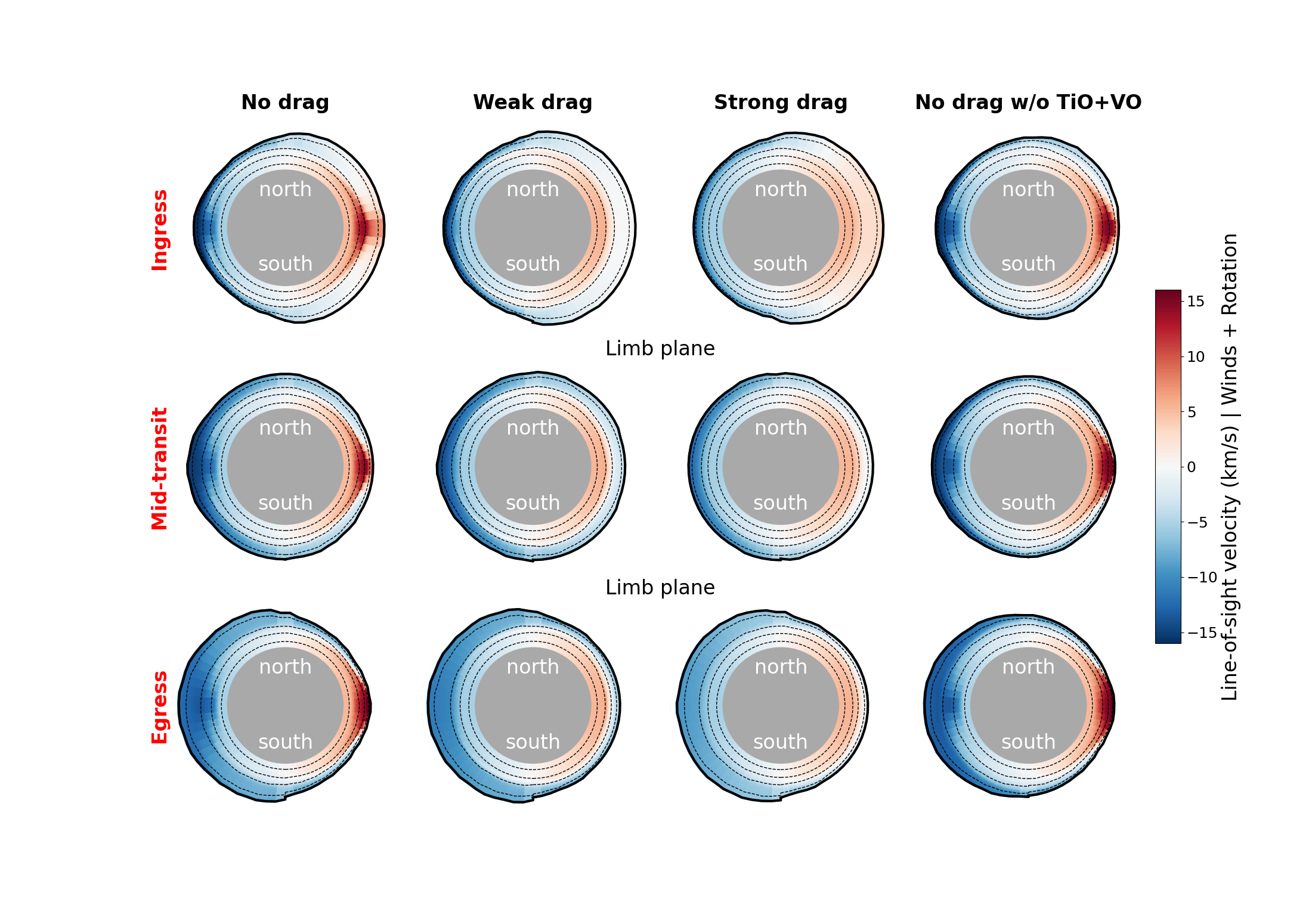}}

\vspace{-41pt}

\caption{Same as Figure \ref{fig:app_c1}, but now for the line-of-sight velocity due to winds and planetary rotation.}
\label{fig:app_c4}
\end{figure*}


\bsp	
\label{lastpage}
\end{document}